\documentclass{article}

\usepackage[letterpaper,top=2cm,bottom=2cm,left=3cm,right=3cm,marginparwidth=1.75cm]{geometry}

\usepackage[colorlinks=true, allcolors=blue]{hyperref}

\usepackage{lineno}
\usepackage{xcolor}
\usepackage{graphicx}% Include figure files
\usepackage{dcolumn}% Align table columns on decimal point
\usepackage{bm}% bold math
\usepackage{amsmath}
\usepackage{mathrsfs} %花体
\usepackage{calligra}
\usepackage[T1]{fontenc}
\usepackage{upgreek}
\usepackage{appendix}
\usepackage[utf8]{inputenc}

\title{Multiple scattering and diffusion of scalar coherent waves in a group of small spheroidal particles with random orientations}
\author{
	Mingyuan Ren, Yajing Qiao, Ning Zhou, Jianrui Gong, Yang Zhou, Yu Zhang * \\
	Department of Physics, Harbin Institute of Technology, \\
	No.92, XiDazhi Street, Harbin 150001, China \\
	\texttt{\{Mingyuan Ren\} hitphyrmy@126.com} \\
	\texttt{\{Yu Zhang\} zhangyuhitphy@163.com} 
}

\begin{document}
\maketitle

\begin{abstract}
In this manuscript we study multiple scattering and diffusion of scalar wave in a group of monodisperse spheroidal particles with random orientations. We begin by fixing a spheroid in a prolate spheroidal coordinate system, and attain the expansion of the scalar Green's function in this space. The expansion is firstly based on spheroidal wave functions, and then we transform it into the expansion of spherical wave functions. Next, we average the Green's function over the orientations of the spheroid to get the averaged transition operator. Finally, we calculate the transport mean free path and anisotropy factor for the spheroidal particles group, based on the irreducible vertex in the Bethe-Salpeter equation. The approaches to get the average transition operator and the mean free paths in this manuscript will be of benefit to the research area of multiple scattering by non-spherical particles.
\end{abstract}

\section{\label{sec:level1}Introduction}

One of the most widely investigated problems in waves scattering is spherical particles scattering, which is initially theoretically solved by Gustav Mie \cite{Bohren2008}. Mie theory has been used in many applications since spherical particle is an efficient model in numerical calculations \cite{Wiscombe:80}. However, there are more particles in natural environment that is non-spherical, including spheroids, cylinders, and even particles aggregates \cite{Mishchenko2000}. Accordingly, many researchers have proposed various kinds of approaches to obtain scattered fields by non-spherical particles. Such approaches include Waterman's T-matrix method \cite{Mishchenko2002}, separation of variables method \cite{Mishchenko2014,Farafonov2013}, dyadic Green's function method \cite{Tai_1971,LiLewei2004}, etc.      These three methods all concern eigenfunction expansions in corresponding coordinate system and provide exact solutions, but they are applicable to different theoretical framework. For example, the Green's function method is the most suitable one to multiple scattering theory.

When a wave propagates in a group of particles, it is possibly scattered many times by different particles. This process will gradually become dominant as the propagation distance exceeds the scattering mean free path \cite{Akkermans2007,Carminati2021,Mishchenko2006}. The single-scattering methods mentioned above are not enough in this case, and multiple scattering theory should be introduced.
For matter waves and electrons, the quantum kinetic equation in multiple scattering processes has been derived and further used to analyze weak localization \cite{Piraud_2013,Vollhardt1980}. The Diffuson and the Cooperon are proved to have close connection to the irreducible vertex.
Recently, multiple scattering of polarized light in point scatterers, as well as that in disordered media exhibiting short-range structural correlations are theoretically derived, based on the Bethe-Salpeter equation \cite{Tiggelen_2016,Carminati2014}.  The latter paper has used Henyey-Greenstein phase function to replace the irreducible vertex, and discussed depolarization effect of different eigenmodes. Thus, the irreducible vertex plays a central role in multiple scattering, but in fact, it is determined by each single scattering process in the independent scattering approximation \cite{Amic_1996}.

If the particles are sparsely distributed in the medium, we can assume that any two of them have no correlations and this is called independent scattering approximation (ISA). In this approximation, the irreducible vertex can be directly calculated by the transition operator \cite{Sheng2006,RevModPhys.71.313,Carminati2021}, while this operator can be directly calculated from the Green's function for one single scatterer \cite{Tsang1980,Tsang1981}. Tai has given the dyadic Green's function for a spherical object and several kinds of non-spherical objects \cite{Tai_1971}, and Li et al. have derived the dyadic Green's function for a spheroid \cite{LiLewei2001,LiLewei2004}. Based on these Green' functions, the irreducible vertex and diffusion constant can be calculated, which gives results of diffusion and weak localization.

However, the Green's functions in \cite{Tai_1971,LiLewei2004} are calculated in the case that the scatterer has a fixed orientation, which is not the case of natural environment. If one wants to derive multiple scattering in a group of non-spherical particles, the averaging over the orientations of particles is necessary. Using Waterman's T-matrix method and the orthogonality of Wigner D function, the averaged scattering amplitude could be calculated, which could further be used to calculate scattering and extinction cross section \cite{Mishchenko2002,Mishchenko2000}. According to this process, we show how to calculate the averaged transition operator in this manuscript, and use this operator to calculate the mean free paths. Compared to T-matrix, transition operator is more suitable to multiple scattering theory. We believe the demonstration of this manuscript will benefit the research area of non-spherical particles scattering, and may have potential applications in dense discrete particles scattering. 

This manuscript is organized as follows. In Section \ref{sec:The scalar transition operator}, we show the relation between scalar scattering Green's function and the corresponding transition operator. Then we calculate the Green's function for a fixed spheroid in Subsection \ref{sec:The scalar Green's function for a fixed spheroid}, and demonstrate how to get the average transition operator in Subsection \ref{sec:Averaging the Green's function and the transition operator}. Next, in Subsection \ref{sec:Transition operator in multiple scattering theory}, we introduce the multiple scattering theory in the formalism of transition operator, and introduce the diffusion approximation and the far-field approximation in Subsection \ref{sec:far field approximation}. Finally, we calculate the scattering and transport mean free paths of monodisperse spheroidal particles group in Subsection \ref{sec:The transport mean free path}. The conclusion is drawn in Section \ref{sec:conclusion}.

\section{The scalar transition operator for a spheroid of random orientation}
\label{sec:The scalar transition operator}

\subsection{From Green's function of a dielectric particle to its transition operator}

A transition operator $T_j$ of a single particle (we use the subscript $j$ to represent the $j$th particle in the particles group) describes how a light field arrives at this particle, possibly bends its direction many times inside this particle, and then leaves it for good.  This process contains a number of terms called Born series, which could be expressed by
\begin{align}
	\label{eq:T = U + U G0 U + U G0 U G0 U + ...}
	T_j \left( \mathbf{r}, \mathbf{r}' \right) =& U\left( \mathbf{r} \right) \delta \left( \mathbf{r}, \mathbf{r}' \right) 
	+ U\left( \mathbf{r} \right) G_0 \left( \mathbf{r}, \mathbf{r}' \right) U\left( \mathbf{r}' \right) 
	\nonumber\\  
	&+ \int \mathrm{d} \mathbf{r}_1 U\left( \mathbf{r} \right) G_0 \left( \mathbf{r}, \mathbf{r}_1 \right) U\left( \mathbf{r}_1 \right) G_0 \left( \mathbf{r}_1, \mathbf{r}' \right) U\left( \mathbf{r}' \right)  
	\nonumber\\
	&+ \cdots, \quad
\end{align}
where 
\begin{eqnarray}
	\label{eq:U(r)}
	U (\mathbf{r}) = 
	\begin{cases}
		k_0^2 \delta \epsilon _p, \quad\,\,\, \mathrm{for}\,\, \mathbf{r}\,\,  \mathrm{inside } \,\, \mathrm{the}\,\, \mathrm{particle},
		\nonumber\\ 
		0, \qquad\quad\,\,  \mathrm{for}\,\, \mathbf{r}\,\, \mathrm{outside } \,\, \mathrm{the}\,\, \mathrm{particle}
	\end{cases}  
\end{eqnarray}
is the scattering potential function of this particle, with $k_0$ the wavenumber in the host medium and $\delta \epsilon_p = \epsilon_p - \epsilon_0$ the difference of dielectric permittivity between the particle $\epsilon_p$ and the host medium $\epsilon_0$.  $G_0$ is the free space Green's function. Noticing the iterative form of Eq. (\ref{eq:T = U + U G0 U + U G0 U G0 U + ...}), we rewrite it as
\begin{align}
	\label{eq:T = U + U G0 T}
	T_j \left( \mathbf{r}, \mathbf{r}' \right) = U\left( \mathbf{r} \right) \delta \left( \mathbf{r}, \mathbf{r}' \right) 
	+   \int \mathrm{d} \mathbf{r}_1    T_j \left( \mathbf{r}, \mathbf{r}_1 \right) G_0 \left( \mathbf{r}_1, \mathbf{r}' \right) U\left( \mathbf{r}' \right).
\end{align}

Those terms in Eq. (\ref{eq:T = U + U G0 U + U G0 U G0 U + ...}) could certainly be calculated one by one, but it is difficult and also unnecessary to perform this calculation. Instead, we can use the technique of Green's function to get the expression of $T_j$. Now, we firstly show the relationship between $T_j$ and its corresponding Green's function. 
Solving the scalar Helmholtz equation
\begin{eqnarray}
	\label{eq:Helmholtz equation}
	\nabla^2 G \left( \mathbf{r}, \mathbf{r}' \right) +  k_0^2  G \left( \mathbf{r}, \mathbf{r}' \right) = -\delta  \left( \mathbf{r}, \mathbf{r}' \right), 
\end{eqnarray}
we get the expression of the Green's function 
\begin{align}
	\label{eq:G = G0 + G0 U Gs}
	G_e\left( \mathbf{r}, \mathbf{r}'  \right) =  G_0\left( \mathbf{r}, \mathbf{r}'  \right) 
	+ \int \mathrm{d} \mathbf{r}_1  G_0\left( \mathbf{r}, \mathbf{r}_1 \right) U\left( \mathbf{r}_1 \right) G_e\left( \mathbf{r}_1, \mathbf{r}' \right),
\end{align}
where we follow Ref. \cite{Tai_1971} to write the Green's function as $G_e$. By noticing the iterative form of Eq. (\ref{eq:G = G0 + G0 U Gs}) and comparing it with Eq. (\ref{eq:T = U + U G0 T}), we get another expression of the Green's function
\begin{align}
	\label{eq:G = G0 + G0 T G0}
	G_e\left( \mathbf{r}, \mathbf{r}'  \right) =  G_0\left( \mathbf{r}, \mathbf{r}'  \right) 
	+ \int \mathrm{d} \mathbf{r}_2 \mathrm{d} \mathbf{r}_1  G_0\left( \mathbf{r}, \mathbf{r}_2 \right) T_j \left( \mathbf{r}_2, \mathbf{r}_1 \right) G_0\left( \mathbf{r}_1, \mathbf{r}' \right),
\end{align}
Multiplying Eq. (\ref{eq:T = U + U G0 T}) by $G_0 \left( \mathbf{r}'', \mathbf{r} \right) $ and performing integration over $\mathbf{r}$, we get
\begin{align}
	\label{eq:a middle equation}
	\int \mathrm{d} \mathbf{r} G_0 \left( \mathbf{r}'', \mathbf{r} \right)  T_j \left( \mathbf{r}, \mathbf{r}' \right) =  G_0 \left( \mathbf{r}'', \mathbf{r}' \right) U\left( \mathbf{r}' \right)  
	+  \int \mathrm{d} \mathbf{r} \mathrm{d} \mathbf{r}_1  G_0 \left( \mathbf{r}'', \mathbf{r} \right)  T_j \left( \mathbf{r}, \mathbf{r}_1 \right) G_0 \left( \mathbf{r}_1, \mathbf{r}' \right) U\left( \mathbf{r}' \right)  . 
\end{align}
Comparing Eq. (\ref{eq:a middle equation}) with Eq. (\ref{eq:G = G0 + G0 T G0}) we get
\begin{eqnarray}
	\label{eq:Tj with Ge}
	\int \mathrm{d} \mathbf{r} G_0 \left( \mathbf{r}'', \mathbf{r} \right)  T_j \left( \mathbf{r}, \mathbf{r}' \right) =  G_e \left( \mathbf{r}'', \mathbf{r}' \right) U\left( \mathbf{r}' \right)  ,
\end{eqnarray}
whose Fourier transformation is
\begin{align}
	\label{eq:Tj with Ge Fourier domain}
	G_0 \left( \mathbf{p}  \right) T_j \left( \mathbf{p}, \mathbf{p}' \right)    
	=  \int \mathrm{d} \mathbf{r} \mathrm{d} \mathbf{r}'   \mathrm{e} ^{-\mathrm{i} \mathbf{p} \cdot \mathbf{r} }  G_e \left( \mathbf{r}'', \mathbf{r}' \right) U\left( \mathbf{r}' \right)  \mathrm{e} ^{ \mathrm{i} \mathbf{p}' \cdot \mathbf{r}' },
\end{align}
in which 
\begin{eqnarray}
	\label{eq:Tj transformation definition}
	T_j \left( \mathbf{p}, \mathbf{p}' \right) = \int \mathrm{d} \mathbf{r} \mathrm{d} \mathbf{r}'   \mathrm{e} ^{-\mathrm{i} \mathbf{p} \cdot \mathbf{r} }
	T_j \left( \mathbf{r}, \mathbf{r}' \right)
	\mathrm{e} ^{\mathrm{i} \mathbf{p}' \cdot \mathbf{r}' }.  
\end{eqnarray}
Eq. (\ref{eq:Tj with Ge Fourier domain}) will be used to calculate transition operator.

\subsection{The scalar Green's function for a fixed dielectric spheroid}
\label{sec:The scalar Green's function for a fixed spheroid}

Applying the boundary condition to a non-spherical particle in a general coordinate system is not as simple as Mie theories. But fortunately, we have other coordinate systems to choose. Using a prolate spheroidal coordinate system, Li et al. have derived the dyadic Green's function for a dielectric spheroid \cite{LiLewei2001,LiLewei2004}. Following their work, we here develop the scalar Green's function for a dielectric spheroid.

\begin{figure}
	\centering
	\includegraphics[width = 6.0cm]{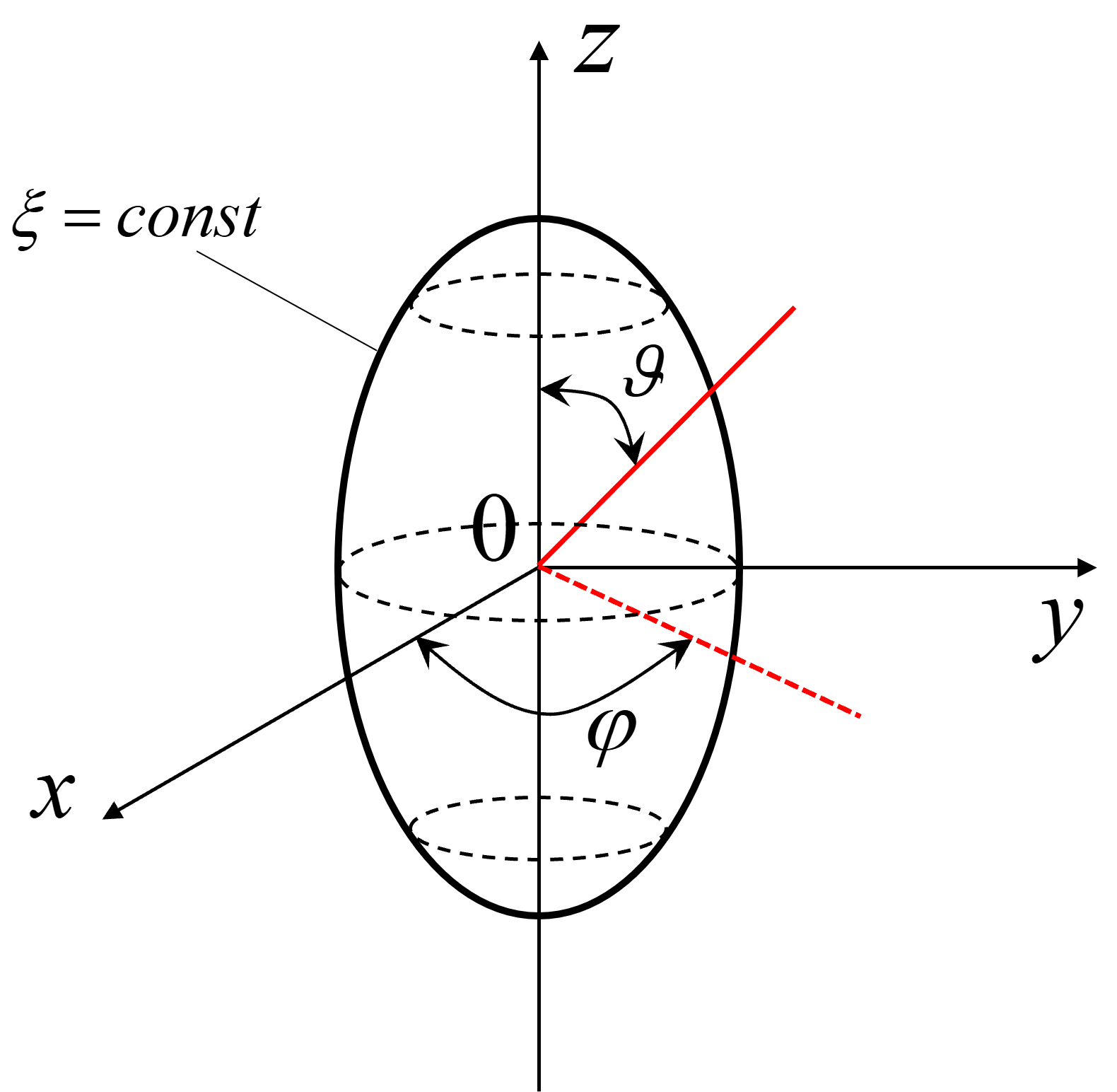}
	\caption{Prolate spheroidal coordinate system $\left( \xi, \vartheta, \varphi \right) $}
	\label{fig:spheroidal_coordinate_system}
\end{figure}

We consider a spheroidal particle centered at the origin of the coordinate system, whose dielectric permittivity is homogeneous. We keep its orientation steady for now, as shown by Figure \ref{fig:spheroidal_coordinate_system}.
We use a prolate spheroidal coordinate system $\left( \xi, \vartheta, \varphi \right) $, which results from rotating the two-dimensional elliptic coordinate system about the focal axis of the ellipse. The relation between the prolate spheroidal coordinates and the Cartesian coordinates are given by
\begin{align}
	\label{eq:relation two coordinate systems}
	x &= a \mathrm{sinh} \xi  \mathrm{sin} \vartheta \mathrm{cos} \varphi ,
	\nonumber\\
	y &= a \mathrm{sinh} \xi  \mathrm{sin} \vartheta \mathrm{sin} \varphi ,
	\nonumber\\
	z &= a \mathrm{cosh} \xi  \mathrm{cos} \vartheta  ,
\end{align}
where $a$ is the semi-interfocal distance of the spheroid. The radial coordinate $\xi$ forms a series of spheroidal surfaces. Now we can express the scalar Helmholtz equation in spheroidal coordinates.
By the usual procedure of the separation of variables, we obtain the scalar wave eigenfunctions 
\begin{eqnarray}
	\label{eq:scalar eigenfunctions}
	\psi_{mn}^{(i)} \left( k_0 a, \mathbf{r} \right)  = R_{mn}^{(i)} \left( k_0 a, \mathrm{cosh} \xi  \right)  \mathscr{P}_n^m  \left( k_0 a, \mathrm{cos} \vartheta  \right) \mathrm{e}^{\mathrm{i} m \varphi}
\end{eqnarray}
where the radial function $R_{mn}^{(i)} $ satisfies Eq. (\ref{eq:R differential equation}) and the angular function $\mathscr{P}_n^m$ satisfies Eq. (\ref{eq:P differential equation}). 
Using $\psi_{mn}^{(i)}$, the free space Green's function could be expressed by \cite{LiLewei2001,LiLewei2004}
\begin{align}
	\label{eq:G0 expansion}
	G_0 \left( \mathbf{r}, \mathbf{r}' \right) 
	&= \frac{\mathrm{i}k_0}{2\uppi} \sum_{n=0}^{\infty} \sum_{m=0}^{n} \left( 2-\delta_{m,0}  \right)    
	\nonumber\\       &\times 
	R_{mn}^{(3)} \left( k_0 a, \mathrm{cosh} \xi^{\ge}  \right)
	R_{mn}^{(1)} \left( k_0 a, \mathrm{cosh} \xi^{\le}  \right)
	\nonumber\\       &\times
	\overline{\mathscr{P}}_n^m  \left( k_0 a, \mathrm{cos} \vartheta  \right) 
	\overline{\mathscr{P}}_n^m  \left( k_0 a, \mathrm{cos} \vartheta'  \right)
	\mathrm{e}^{\mathrm{i} m \left( \varphi - \varphi' \right) }, 
\end{align}
where $\mathbf{r}^{\ge}$, $\mathbf{r}^{\le}$ mean $\xi^{\le} = min \{ \xi, \xi' \}$, $\xi^{\ge} = max \{ \xi, \xi' \}$ , respectively, in spheroidal coordinates. And they mean $r^{\le} = min \{ r, r' \}$, $r^{\ge} = max \{ r, r' \}$ , respectively, in spherical coordinates. When $\xi$ reaches infinity, the radial function $R_{mn}^{(i)}$ tends to be a spherical wave, and when $\xi$ reaches zero, $R_{mn}^{(i)}$ must be finite, so we choose $R_{mn}^{(3)}$ for $\xi^{\ge}$ and $R_{mn}^{(1)}$ for $\xi^{\le}$. The radial function $R_{mn}^{(i)}$ and the normalized angular function $\overline{\mathscr{P}}_n^m$ are shown in Appendix. \ref{sec:spheroidal functions}. 

By the method of scattering superposition, we can construct the Green's function $G_e$ for a dielectric spheroid
\begin{align}
	\label{eq:composition of Ge first}
	G_e^{(22)} \left( \mathbf{r}, \mathbf{r}' \right) &= G_0  \left( \mathbf{r}, \mathbf{r}' \right)  + G_{es}^{(22)} \left( \mathbf{r}, \mathbf{r}' \right), \quad\, \xi\le \xi_0
	\nonumber\\
	G_e^{(12)} \left( \mathbf{r}, \mathbf{r}' \right) &= G_{es}^{(12)} \left( \mathbf{r}, \mathbf{r}' \right) , \qquad\qquad\quad\quad \,  \xi \ge \xi_0,
\end{align}
where $G_e^{(12)}$ means that the source is located in region 1, namely outside the particle, and the receiving point is located in region 2, namely inside the particle. Accordingly, $G_e^{(22)}$ means that the source and the receiving point are both located inside the particle. $G_e^{(21)}$ and $G_e^{(11)}$ are not given here, since we can deduce from Eq. (\ref{eq:Tj with Ge}) that they will vanish in the transition operator because of the presence of $U$. $G_{es}$ is the scattered part in $G_e$. It is important that the free space Green's function here has a wavenumber of $k_s = k_0 \sqrt{\epsilon_p}$ because it describes propagation inside the scatterer. Considering the composition of $G_0$, the scattered part $G_{es}$ must have the form \cite{Tai_1971}
\begin{align}
	\label{eq:Ges first}
	G_{es}^{(22)} \left( \mathbf{r}, \mathbf{r}' \right) 
	&=\frac{\mathrm{i}k_s}{2\uppi} \sum_{n=0}^{\infty} \sum_{m=0}^{n} \left( 2-\delta_{m,0}  \right)     A_{mn} 
	\nonumber\\       &\times 
	R_{mn}^{(1)} \left( k_s a, \mathrm{cosh} \xi   \right)
	R_{mn}^{(1)} \left( k_s a, \mathrm{cosh} \xi'  \right)
	\nonumber\\       &\times
	\overline{\mathscr{P}}_n^m  \left( k_s a, \mathrm{cos} \vartheta  \right) 
	\overline{\mathscr{P}}_n^m  \left( k_s a, \mathrm{cos} \vartheta'  \right)
	\mathrm{e}^{\mathrm{i} m ( \varphi - \varphi' ) },
	\nonumber\\
	G_{es}^{(12)} \left( \mathbf{r}, \mathbf{r}' \right) 
	&=\frac{\mathrm{i}k_s}{2\uppi} \sum_{n=0}^{\infty} \sum_{m=0}^{n} \left( 2-\delta_{m,0}  \right)     C_{mn} 
	\nonumber\\       &\times 
	R_{mn}^{(3)} \left( k_0 a, \mathrm{cosh} \xi   \right)
	R_{mn}^{(1)} \left( k_s a, \mathrm{cosh} \xi'  \right)
	\nonumber\\       &\times
	\overline{\mathscr{P}}_n^m  \left( k_s a, \mathrm{cos} \vartheta  \right) 
	\overline{\mathscr{P}}_n^m  \left( k_s a, \mathrm{cos} \vartheta'  \right)
	\mathrm{e}^{\mathrm{i} m ( \varphi - \varphi' ) }.
\end{align}
The coefficients $A_{mn}$ and $C_{mn}$ are obtained in Appendix \ref{sec:Amn and Cmn}.

\subsection{Averaging the Green's function and the transition operator}
\label{sec:Averaging the Green's function and the transition operator}

In a medium full of spherical particles, the rotation of each sphere has no impact on the transition operator, meaning that the operator is absolutely spherically symmetric. To get the average transition operator $\left\langle T \right\rangle $, we only need to average over the possible positions of all particles. However, this simplicity breaks down if the particle is non-spherical, since the orientations of particles will influence $T_j$, and must be considered in averaging. So we perform the averaging on its orientation for one single spheroid centered at the origin of the coordinate system. In light of the expansion of Green's function, as well as the relationship Eq. (\ref{eq:Tj with Ge Fourier domain}), we realize that it is useful to first calculate $  G_e U  $, which is
\begin{align}
	\label{eq:<Tj>}
	    G_e  \left( \mathbf{r}, \mathbf{r}' \right) U\left( \mathbf{r}' \right) 
	=  k_0^2 \delta \epsilon_p   G^{(22)}_{e}  \left( \mathbf{r}, \mathbf{r}' \right)    + k_0 \delta \epsilon_p   G^{(12)}_{e}  \left( \mathbf{r}, \mathbf{r}' \right)  .
\end{align}
$G^{(12)}_{e}$ and $G^{(22)}_{e}$ vanish because of the existence of $U$.

\begin{figure}[htbp]
	\centering
	\begin{minipage}{0.46\textwidth}
		\centering
		\includegraphics[width=\textwidth]{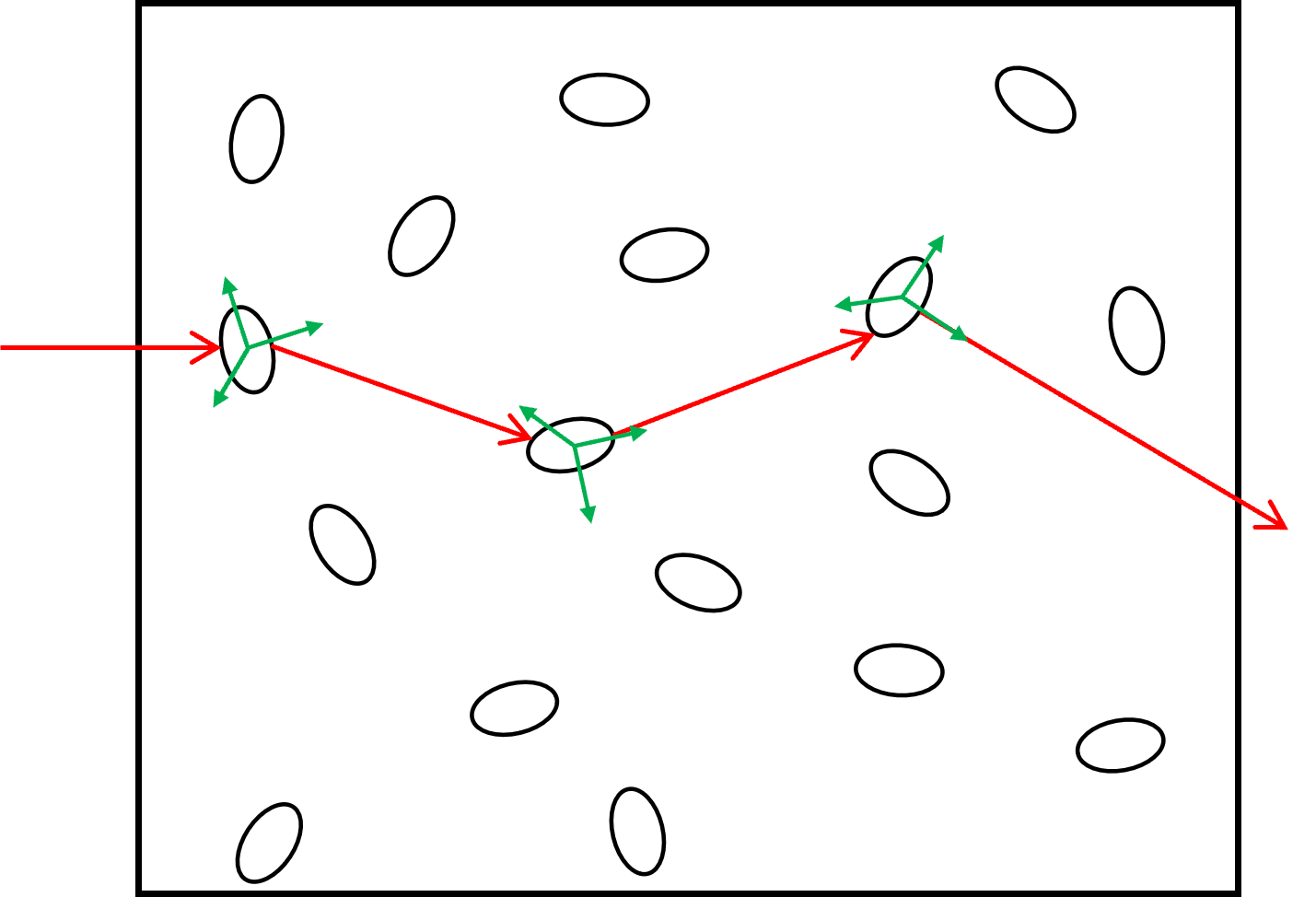}
		\caption{A field (the red solid line) scattering multiply in a group of spheroidal particles.  The local prolate spheroidal coordinate systems are sketched in green for each particle.}
		\label{fig:multiple_scattering_in_spheroidal_particles_group}
	\end{minipage}
	\hspace{0.05\textwidth}
	\centering
	\begin{minipage}{0.46\textwidth}
		\centering
		\includegraphics[width=\textwidth]{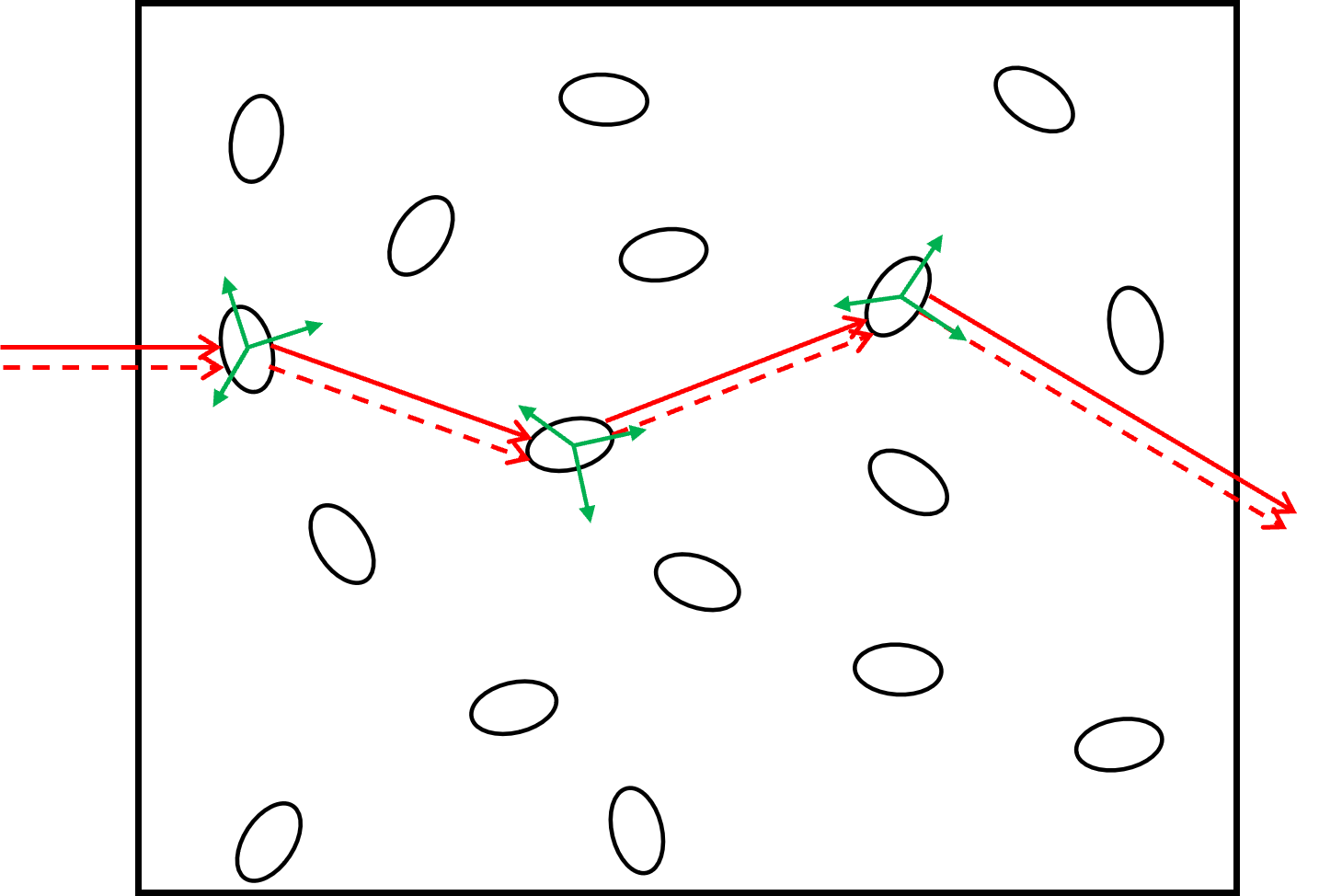}
		\caption{Diffusion of scalar coherent wave in a group of spheroidal particles. The red dashed line represents the conjugate field. The local prolate spheroidal coordinate systems are sketched in green for each particle.}
		\label{fig:diffusion_in_spheroidal_particles_group}
	\end{minipage}
\end{figure}

In the last subsection, we calculate the Green's function by aligning the symmetry-axis of the spheroid with $z$-axis of Cartesian coordinates. However, in natural environment, the orientations of particles are random, as shown by Figure \ref{fig:multiple_scattering_in_spheroidal_particles_group}. We must take it into consideration that the expansions of Green's function in the laboratory coordinate system are different from that in the particle coordinate system. Now, let us assume that the Euler angles $\alpha, \beta, \gamma$ determine the rotation from the laboratory coordinate system $S^{(L)}\{ x^{(L)},y^{(L)},z^{(L)}  \}$ to the particle coordinate system $S^{(P)}\{ x^{(P)},y^{(P)},z^{(P)}  \}$. To average the operator $ G_eU$, we need to perform this averaging over all orientations of particles. This means that, we fix ourselves in $S^{(L)}$, rotate $S^{(P)}$, and take into account all values of $\alpha, \beta, \gamma$. However, we can conversely fix ourselves in $S^{(P)}$, rotate $S^{(L)}$, and average $G_eU$. Obviously the latter manipulation is simpler than the former one.

Before calculating $\left\langle G_e U \right\rangle $, we call for another step. The randomness of the orientation is actually from the perspective of $(\theta, \phi)$, namely the spherical coordinates, and the angles $\vartheta$ of spheroidal coordinates are not homogeneous in the natural space. Therefore, it is better to transform those functions into spherical coordinates, if we need to average over Euler angles. First, by the relation between the spheroidal wave function and the spherical wave function \cite{Farafonov_2016}
\begin{align}
	R_{mn}^{(i)} & \left( k_0a, \xi \right)  \overline{\mathscr{P}}_n^m \left( k_0a, \mathrm{cos} \vartheta \right)  
	=
	\sum_{l=m}^{\infty}  \mathrm{i}^{l-n} d_{l-m}^{mn} (k_0a)
	\frac{N_{mn}(0)}{N_{ml}(k_0a)}
	z^{(i)}_l (k_0r) \overline{P}_l^m\left( \mathrm{cos} \theta \right)  ,
\end{align}
where $z^{(i)}_l(r)$ is the spherical Bessel function of order $l$ and type $i$, and $\overline{P}_l^m \left( \mathrm{cos} \theta \right) = N^{-1}_{mn}(0) P_l^m \left( \mathrm{cos} \theta \right) $ is the associated Legendre function with the normalizing factor given by Eq. (\ref{eq:Nmn(0)}). And $P_l^m \left( \mathrm{cos} \theta \right)$ is the associated Legendre functions with the general definition. Then the scattering Green's function could be expressed by
\begin{align}
	\label{eq:Ges(22) second}
	G_{es}^{(22)} \left( \mathbf{r}, \mathbf{r}' \right) 
	&=\frac{\mathrm{i}k_s}{2\uppi} \sum_{n=0}^{\infty} \sum_{m=0}^{n} \left( 2-\delta_{m,0}  \right)     A_{mn} \mathrm{e}^{\mathrm{i} m ( \phi - \phi' ) }
	\nonumber\\ &  \times
	\sum_{l=m}^{\infty}  \mathrm{i}^{l-n} d_{l-m}^{mn} (k_sa)
	\frac{N_{mn}(0)}{N_{ml}(k_sa)}
	z^{(1)}_l (k_sr) \overline{P}_l^m\left( \mathrm{cos} \theta \right)
	\nonumber\\& \times
	\sum_{l'=m}^{\infty}  \mathrm{i}^{l'-n} d_{l'-m}^{mn} (k_sa)
	\frac{N_{mn}(0)}{N_{ml'}(k_sa)}
	z^{(1)}_{l'} (k_sr') \overline{P}_{l'}^m\left( \mathrm{cos} \theta' \right) ,
\end{align}
\begin{align}
	\label{eq:Ges(12) second}
	G_{es}^{(12)} \left( \mathbf{r}, \mathbf{r}' \right) 
	&=\frac{\mathrm{i}k_s}{2\uppi} \sum_{n=0}^{\infty} \sum_{m=0}^{n} \left( 2-\delta_{m,0}  \right)     C_{mn} \mathrm{e}^{\mathrm{i} m ( \phi - \phi' ) }
	\nonumber\\ & \times
	\sum_{l=m}^{\infty}  \mathrm{i}^{l-n} d_{l-m}^{mn} (k_0a)
	\frac{N_{mn}(0)}{N_{ml}(k_0a)}
	z^{(3)}_l (k_0r) \overline{P}_l^m\left( \mathrm{cos} \theta \right)
	\nonumber\\& \times
	\sum_{l'=m}^{\infty}  \mathrm{i}^{l'-n} d_{l'-m}^{mn} (k_sa)
	\frac{N_{mn}(0)}{N_{ml'}(k_sa)}
	z^{(1)}_{l'} (k_sr') \overline{P}_{l'}^m\left( \mathrm{cos} \theta' \right) .
\end{align}

The Wigner D function $D_{m'm}^l \left( \alpha, \beta, \gamma \right) $ can be used to calculate the rotation from $S^{(L)}  \{ r, \theta^{(L)}, \phi^{(L)} \}$ to $S^{(P)}  \{ r, \theta^{(P)}, \phi^{(P)} \}$ \cite{Varshalovich1988}
\begin{align}
	Y_l^m \left( \theta^{(P)}, \phi^{(P)} \right)  
	= \sum_{k=-l}^{l}  Y_l^{k} \left( \theta^{(L)}, \phi^{(L)} \right)  D_{km}^l \left( \alpha, \beta, \gamma \right) ,
\end{align}
where $\sqrt{2\uppi} \, Y_l^m (\theta, \phi) =  \overline{P}_l^m (\mathrm{cos} \theta) \mathrm{e}^{\mathrm{i} m \phi }$ is the spherical harmonics. It is important to note that the angular coordinate $\varphi$ in a spheroidal coordinate system is identical to the angular coordinate $\phi$ in its corresponding spherical coordinate system \cite{Farafonov_2016}. Averaging the scattering Green's function Eq. (\ref{eq:Ges(22) second}) and (\ref{eq:Ges(12) second}) yields
\begin{align}
	\label{eq:<Ges(22)> first}
	\left\langle G_{es}^{(22)} \left( \mathbf{r}^{(P)}, \mathbf{r}'^{(P)} \right) \right\rangle &
	=\frac{\mathrm{i}k_s}{2\uppi} \sum_{n=0}^{\infty} \sum_{m=0}^{n} \left( 2-\delta_{m,0}  \right)     A_{mn} 
	\nonumber\\ & \quad\times
	\sum_{l=m}^{\infty}  \mathrm{i}^{l-n} d_{l-m}^{mn} (k_sa)
	\frac{N_{mn}(0)}{N_{ml}(k_sa)}
	z^{(1)}_l (k_sr) 
	\sum_{l'=m}^{\infty}  \mathrm{i}^{l'-n} d_{l'-m}^{mn} (k_sa)
	\frac{N_{mn}(0)}{N_{ml'}(k_sa)}
	z^{(1)}_{l'} (k_sr') 
	\nonumber\\ &\quad\times  2\uppi
	\sum_{k=-l}^{l}  Y_l^{k} \left( \theta^{(L)}, \phi^{(L)} \right)  
	\sum_{k'=-l'}^{l'}  {Y_{l'}^{k'}}^* \left( \theta^{'(L)}, \phi^{'(L)} \right) 
	\left\langle D_{km}^l \left( \alpha, \beta, \gamma \right) 
	{D_{k'm}^{l'} }^* \left( \alpha, \beta, \gamma \right) \right\rangle 
	\nonumber\\&
	=\frac{\mathrm{i}k_s}{4\uppi} \sum_{n=0}^{\infty} \sum_{m=0}^{n} \left( 2-\delta_{m,0}  \right)     A_{mn} 
	\sum_{l=m}^{\infty} \left[  \mathrm{i}^{l-n} d_{l-m}^{mn} (k_sa)
	\frac{N_{mn}(0)}{N_{ml}(k_sa)} \right] ^2
	\nonumber\\ &\quad\times
	z^{(1)}_l (k_sr)     z^{(1)}_{l} (k_sr') 
	P_l(\mathrm{cos} \Theta) ,
\end{align}

\begin{align}
	\label{eq:<Ges(12)> first}
	\left\langle G_{es}^{(12)} \left( \mathbf{r}^{(P)}, \mathbf{r}'^{(P)} \right) \right\rangle 
	&=\frac{\mathrm{i}k_s}{2\uppi} \sum_{n=0}^{\infty} \sum_{m=0}^{n} \left( 2-\delta_{m,0}  \right)     C_{mn}  
	\nonumber\\ & \quad\times
	\sum_{l=m}^{\infty}  \mathrm{i}^{l-n} d_{l-m}^{mn} (k_0a)
	\frac{N_{mn}(0)}{N_{ml}(k_0a)}
	z^{(3)}_l (k_0r)  
	\sum_{l'=m}^{\infty}  \mathrm{i}^{l'-n} d_{l'-m}^{mn} (k_sa)
	\frac{N_{mn}(0)}{N_{ml'}(k_sa)}
	z^{(1)}_{l'} (k_sr') 
	\nonumber\\ &\quad\times   2\uppi
	\sum_{k=-l}^{l}  Y_l^{k} \left( \theta^{(L)}, \phi^{(L)} \right)  
	\sum_{k'=-l'}^{l'}  {Y_{l'}^{k'}}^* \left( \theta^{'(L)}, \phi^{'(L)} \right) 
	\left\langle D_{km}^l \left( \alpha, \beta, \gamma \right) 
	{D_{k'm}^{l'} }^* \left( \alpha, \beta, \gamma \right) \right\rangle 
	\nonumber\\&
	=\frac{\mathrm{i}k_s}{4\uppi} \sum_{n=0}^{\infty} \sum_{m=0}^{n} \left( 2-\delta_{m,0}  \right)     C_{mn}  
	\sum_{l=m}^{\infty}  
	\frac{\left[ \mathrm{i}^{l-n}  N_{mn}(0) \right]^2 }{N_{ml}(k_0a) N_{ml}(k_sa)} 
	d_{l-m}^{mn} (k_0a) d_{l-m}^{mn} (k_sa)
	\nonumber\\ &  \quad\times
	z^{(3)}_l (k_0r) 
	z^{(1)}_{l} (k_sr') 
	P_l(\mathrm{cos} \Theta).
\end{align}
where we have used Eq. (\ref{eq:<D*D> calculation}) and Eq. (\ref{eq:the addition theorem of Legendre functions 3rd version}). It is worth noting that $\left\langle G^{(22)}_{es} \right\rangle $ and $\left\langle G^{(12)}_{es} \right\rangle $ are independent of the exact directions $\hat{\mathbf{r}}$ and $\hat{\mathbf{r}'}$, but only dependent on the angle between them, namely $\Theta$. This leads to a very important result, that $\left\langle UG_e \right\rangle $ is independent of the coordinate system we choose. Therefore, Eq. (\ref{eq:<Ges(22)> first}) and (\ref{eq:<Ges(12)> first}) has already given the averaging of one transition operator for a spheroidal particle centered at the origin and with random orientation.
The averaging has no impact on the free space Green's function. It can still be calculated using Eq. (\ref{eq:G0 appendix expansion 2}) or $G_0^{-1}(\mathbf{p}) = p^2 - k_0^2$.

Using Eq. (\ref{eq:appendix formula 1}) and (\ref{eq:appendix formula 2}), Eq. (\ref{eq:<Ges(22)> first}) and (\ref{eq:<Ges(12)> first}) could be Fourier transformed into 
\begin{align}
	\label{eq:<Ges(22)> Fourier}
	\left\langle G_{es}^{(22)} \left( \mathbf{p}, \mathbf{p}' \right) \right\rangle 
	&=\frac{\mathrm{i}k_s}{4\uppi} \sum_{n=0}^{\infty} \sum_{m=0}^{n} \left( 2-\delta_{m,0}  \right)     A_{mn} 
	\sum_{l=m}^{\infty} \left[  \mathrm{i}^{l-n} d_{l-m}^{mn} (k_sa)
	\frac{N_{mn}(0)}{N_{ml}(k_sa)} \right] ^2
	P_l(\mathrm{cos} \Theta_p) 
	\nonumber\\& \times
	4\uppi \mathrm{i}^{n}  
	\frac{pj_n(k_sA)  	j'_n(pA) -  k_s  j'_n(k_sA)  	j_n(pa)}{k_s^2 - p^2} A^2
	\nonumber\\& \times
	4\uppi \mathrm{i}^{n}  
	\frac{p'j_n(k_sA)  	j'_n(p'A) -  k_s  j'_n(k_sA)  	j_n(p'A)}{k_s^2 - p^{'2}} A^2,
\end{align}

\begin{align}
	\label{eq:<Ges(12)> Fourier}
	\left\langle G_{es}^{(12)} \left( \mathbf{p}, \mathbf{p}' \right) \right\rangle 
	&=\frac{\mathrm{i}k_s}{4\uppi} \sum_{n=0}^{\infty} \sum_{m=0}^{n} \left( 2-\delta_{m,0}  \right)     C_{mn}  
	\sum_{l=m}^{\infty}  
	\frac{\left[ \mathrm{i}^{l-n}  N_{mn}(0) \right]^2 }{N_{ml}(k_0a) N_{ml}(k_sa)} 
	d_{l-m}^{mn} (k_0a) d_{l-m}^{mn} (k_sa)
	P_l(\mathrm{cos} \Theta_p) 
	\nonumber\\& \times
	4\uppi \mathrm{i}^{n}  \left[   \frac{\uppi}{2k_0^2} \delta (k_0-p)  - 
	\frac{ph^{(1)}_n(k_0A)  	j'_n(pA) -  k_0  h^{(1)'}_n(k_0A)  	j_n(pA)}{k_0^2 - p^2} A^2  \right] 
	\nonumber\\& \times
	4\uppi \mathrm{i}^{n}  
	\frac{p'j_n(k_sA)  	j'_n(p'A) -  k_s  j'_n(k_sA)  	j_n(p'A)}{k_s^2 - p^{'2}} A^2,
\end{align}
where $A$ is the half length of the major axis of the prolate spheroid, and the angle $\Theta_p$ is the angle between $(\theta_p, \phi_p)$ and $(\theta'_p, \phi'_p)$, given by
\begin{align}
	\mathrm{cos} \Theta_p  = \mathrm{cos} \theta_p  \mathrm{cos} \theta'_p  +  \mathrm{sin} \theta_p  \mathrm{sin} \theta'_p \mathrm{cos} \left( \phi_p - \phi'_p  \right).
\end{align}

 Using Eq. (\ref{eq:Tj with Ge Fourier domain}) and (\ref{eq:<Tj>}) we get
\begin{align}
	\label{eq:Tj relation UGe}
	T_j \left( \mathbf{p}, \mathbf{p}' \right)  
	= k_0^2 \delta \epsilon_p   G_0^{-1} \left( \mathbf{p}  \right)   
	\left[    G^{(22)}_{e}  \left( \mathbf{p}, \mathbf{p}' \right)     +     G^{(12)}_{e}  \left( \mathbf{p}, \mathbf{p}' \right)  \right] .
\end{align}
Multiplying Eq. (\ref{eq:<Ges(22)> Fourier}) and (\ref{eq:<Ges(12)> Fourier}) by $G_0^{-1}(\mathbf{p}) = p^2 - k_0^2$ and setting $p = k_0$ (we will show  in Subsection \ref{sec:far field approximation} the reason why we set this), we get from Eq. (\ref{eq:Tj relation UGe}) that
\begin{align}
	\label{eq:only G(12) remains}
	\left\langle T_j\left( k_0\hat{\mathbf{p}}, k_0\hat{\mathbf{p}'} \right)\right\rangle  &  
	= k_0^2 \delta \epsilon_p   G_0^{-1} \left( k_0\hat{\mathbf{p} }  \right)  
	\left[  \left\langle G^{(22)}_{e}  \left( k_0\hat{\mathbf{p}}, k_0\hat{\mathbf{p}'} \right) \right\rangle   +    \left\langle G^{(12)}_{e}  \left( k_0\hat{\mathbf{p}}, k_0\hat{\mathbf{p}'} \right) \right\rangle \right] 
	\nonumber\\ & 
	= k_0^2 \delta \epsilon_p   G_0^{-1} \left( k_0\hat{\mathbf{p}}  \right) 
	\left\langle G^{(12)}_{es}  \left( k_0\hat{\mathbf{p}}, k_0\hat{\mathbf{p}'} \right) \right\rangle,
\end{align}
where we have noticed that only $\left\langle G^{(12)}_{es} \left( \mathbf{p}, \mathbf{p}' \right)  \right\rangle $ has a term $p^2 - k_0^2$ in its denominator, and the function $G_0$ in Eq. (\ref{eq:composition of Ge first}) has a wavenumber $k_s$. Then the final result of the transition operator is
\begin{align}
	\label{eq:final result of the transition operator}
	\left\langle T_j\left( k_0\hat{\mathbf{p}}, k_0\hat{\mathbf{p}'} \right)  \right\rangle  
	&= 4\uppi   \delta \epsilon_p k_0 k_s  \sum_{n=0}^{\infty} \sum_{m=0}^{n} \left( 2-\delta_{m,0}  \right)     C_{mn}  
	\nonumber\\&\times
	\sum_{l=m}^{\infty} (-1)^{l} N^2_{mn}(0) 
	\frac{ d_{l-m}^{mn} (k_0a) d_{l-m}^{mn} (k_sa) } {N_{ml}(k_0a) N_{ml}(k_sa)} 
	P_l(\mathrm{cos} \Theta_p) 
	\nonumber\\& \times   
	\frac{k_0 j_n(k_sA)  	j'_n(k_0A) -  k_s  j'_n(k_sA)  	j_n(k_0A)}{k_s^2 - k_0^{2}} A^2,
\end{align}
where we have used the fact that \cite{WangZhuxi}
\begin{align}
	j_n(x)h_n^{(1)'} (x) - j'_n(x) h_n^{(1)} (x) = \frac{\mathrm{i}}{x^2}  .
\end{align}

\vspace{1cm}

\subsection{The product of a transition operator and its conjugate}
\label{sec:Averaging the product of a transition operator and its conjugate}

After calculating $\left\langle T_j \right\rangle $, we notice that the product of a transition operator and its conjugate is also of vital importance. We will see in Subsection \ref{sec:Transition operator in multiple scattering theory} that, this product plays a central role in the two-particle Green's function (the intensity propagator). In this subsection we will give the expression of this product, but the calculation of its averaging will be left to Subsection \ref{sec:The transport mean free path}.

Let us now derive the transition operator in the reciprocal space in a spherical coordinate system (without being averaged over the Euler angles). To begin with, we calculate the scattering Green's function in the reciprocal space
\begin{align}
	\label{eq:Ge(12) Fourier}
	G_{es}^{(12)} \left( \mathbf{p}, \mathbf{p}' \right) 
	&=\frac{\mathrm{i}k_s}{2\uppi} \sum_{n=0}^{\infty} \sum_{m=0}^{n} \left( 2-\delta_{m,0}  \right)     C_{mn} \mathrm{e}^{\mathrm{i} m ( \phi_p - \phi'_p ) }
	\nonumber\\ & \times
	\sum_{l_1=m}^{\infty}  \mathrm{i}^{l_1-n} d_{l_1-m}^{mn} (k_0a)
	\frac{N_{mn}(0)}{N_{ml_1}(k_0a)}
	  \overline{P}_{l_1}^m\left( \mathrm{cos} \theta_p \right)
	\sum_{l_2=m}^{\infty}  \mathrm{i}^{l_2-n} d_{l_2-m}^{mn} (k_sa)
	\frac{N_{mn}(0)}{N_{ml_2}(k_sa)}
	 \overline{P}_{l_2}^m\left( \mathrm{cos} \theta'_p \right)
	\nonumber\\& \times
	4\uppi \mathrm{i}^{n}  \left[   \frac{\uppi}{2k_0^2} \delta (k_0-p)  - 
	\frac{ph^{(1)}_n(k_0A)  	j'_n(pA) -  k_0  h^{(1)'}_n(k_0A)  	j_n(pA)}{k_0^2 - p^2} A^2  \right] 
	\nonumber\\& \times
	4\uppi \mathrm{i}^{n}  
	\frac{p'j_n(k_sA)  	j'_n(p'A) -  k_s  j'_n(k_sA)  	j_n(p'A)}{k_s^2 - p^{'2}} A^2 ,
\end{align}
as we have done in the calculation of Eq. (\ref{eq:<Ges(22)> Fourier}).
With the previous experience of Eq. (\ref{eq:only G(12) remains}), we here avoid to calculate $G_{es}^{(22)}$ since it will vanish in the transition operator in the far-field approximation. Now we multiply Eq. (\ref{eq:Ge(12) Fourier}) by its conjugate to get
\begin{align}
	\label{eq:G(12) square}
	\left| G_{es}^{(12)} \right.  & \left.  \left( \mathbf{p}, \mathbf{p}' \right) \right| ^2
	=\frac{\mathrm{i}k_s}{2\uppi} \sum_{n=0}^{\infty} \sum_{m=0}^{n} \left( 2-\delta_{m,0}  \right)     C_{mn} \mathrm{e}^{\mathrm{i} m ( \phi_p - \phi'_p ) }
	\nonumber\\ & \times
	\sum_{l_1=m}^{\infty}  \mathrm{i}^{l_1-n} d_{l_1-m}^{mn} (k_0a)
	\frac{N_{mn}(0)}{N_{ml_1}(k_0a)}
	\overline{P}_{l_1}^m\left( \mathrm{cos} \theta_p \right)
	\nonumber\\&\times
	\sum_{l_2=m}^{\infty}  \mathrm{i}^{l_2-n} d_{l_2-m}^{mn} (k_sa)
	\frac{N_{mn}(0)}{N_{ml_2}(k_sa)}
	\overline{P}_{l_2}^m\left( \mathrm{cos} \theta'_p \right)
	\nonumber\\& \times
	4\uppi \mathrm{i}^{n}  \left[   \frac{\uppi}{2k_0^2} \delta (k_0-p)  - 
	\frac{ph^{(1)}_n(k_0A)  	j'_n(pA) -  k_0  h^{(1)'}_n(k_0A)  	j_n(pA)}{k_0^2 - p^2} A^2  \right] 
	\nonumber\\& \times
	4\uppi \mathrm{i}^{n}  
	\frac{p'j_n(k_sA)  	j'_n(p'A) -  k_s  j'_n(k_sA)  	j_n(p'A)}{k_s^2 - p^{'2}} A^2 
				\nonumber\\ & \qquad\times
	\frac{-\mathrm{i}k_s}{2\uppi} \sum_{n'=0}^{\infty} \sum_{m'=0}^{n'} \left( 2-\delta_{m',0}  \right)     C^*_{m'n'} \mathrm{e}^{-\mathrm{i} m' ( \phi_p - \phi'_p ) }
	\nonumber\\ & \times
	\sum_{l_3=m'}^{\infty}  (-\mathrm{i})^{l_3-n'} d_{l_3-m'}^{m'n'} (k_0a)
	\frac{N_{m'n'}(0)}{N_{m'l_3}(k_0a)}
	\overline{P}_{l_3}^{m'}\left( \mathrm{cos} \theta_p \right)
	\nonumber\\&\times
	\sum_{l_4=m'}^{\infty}  (-\mathrm{i})^{l_4-n'} d_{l_4-m'}^{m'n'} (k_sa)
	\frac{N_{m'n'}(0)}{N_{m'l_4}(k_sa)}
	\overline{P}_{l_4}^{m'}\left( \mathrm{cos} \theta'_p \right)
	\nonumber\\& \times
	4\uppi (-\mathrm{i})^{n'}  \left[   \frac{\uppi}{2k_0^2} \delta (k_0-p)  - 
	\frac{ph^{(1)}_{n'}(k_0A)  	j'_{n'}(pA) -  k_0  h^{(1)'}_{n'}(k_0A)  	j_{n'}(pA)}{k_0^2 - p^2} A^2  \right] 
	\nonumber\\& \times
	4\uppi (-\mathrm{i})^{n'}  
	\frac{p'j_{n'}(k_sA)  	j'_{n'}(p'A) -  k_s  j'_{n'}(k_sA)  	j_{n'}(p'A)}{k_s^2 - p^{'2}} A^2 .
\end{align}
Then we get the transition operator
\begin{align}
	\label{eq:Tj*Tj the Fourier}
	\left|  T_j \right.  &\left. \left( k_0\hat{\mathbf{p}}, k_0\hat{\mathbf{p}'} \right)\right| ^2  = k_0^4 (\delta \epsilon_p)^2  
	\left| G^{(12)}_{es}  \left( k_0\hat{\mathbf{p}}, k_0\hat{\mathbf{p}'} \right) \right|^2
	\nonumber\\&
	= k_0^4 (\delta \epsilon_p)^2     
	\frac{\mathrm{i}k_s}{2\uppi} \sum_{n=0}^{\infty} \sum_{m=0}^{n} \left( 2-\delta_{m,0}  \right)     C_{mn} \mathrm{e}^{\mathrm{i} m ( \phi_p - \phi'_p ) }
	\nonumber\\ & \times
	\sum_{l_1=m}^{\infty}  \mathrm{i}^{l_1-n} d_{l_1-m}^{mn} (k_0a)
	\frac{N_{mn}(0)}{N_{ml_1}(k_0a)}
	\overline{P}_{l_1}^m\left( \mathrm{cos} \theta_p \right)
	\sum_{l_2=m}^{\infty}  \mathrm{i}^{l_2-n} d_{l_2-m}^{mn} (k_sa)
	\frac{N_{mn}(0)}{N_{ml_2}(k_sa)}
	\overline{P}_{l_2}^m\left( \mathrm{cos} \theta'_p \right)
	\nonumber\\& \times
	4\uppi \mathrm{i}^{n}  \frac{\mathrm{i}}{k_0}
	\times
	4\uppi \mathrm{i}^{n}  
	\frac{p'j_n(k_sA)  	j'_n(k_0A) -  k_s  j'_n(k_sA)  	j_n(k_0A)}{k_s^2 - k_0^2} A^2 
	\nonumber\\ & \times
	\frac{-\mathrm{i}k_s}{2\uppi} \sum_{n'=0}^{\infty} \sum_{m'=0}^{n'} \left( 2-\delta_{m',0}  \right)     C^*_{m'n'} \mathrm{e}^{-\mathrm{i} m' ( \phi_p - \phi'_p ) }
	\nonumber\\ & \times
	\sum_{l_3=m'}^{\infty}  (-\mathrm{i})^{l_3-n'} d_{l_3-m'}^{m'n'} (k_0a)
	\frac{N_{m'n'}(0)}{N_{m'l_3}(k_0a)}
	\overline{P}_{l_3}^{m'}\left( \mathrm{cos} \theta_p \right)
	\sum_{l_4=m'}^{\infty}  (-\mathrm{i})^{l_4-n'} d_{l_4-m'}^{m'n'} (k_sa)
	\frac{N_{m'n'}(0)}{N_{m'l_4}(k_sa)}
	\overline{P}_{l_4}^{m'}\left( \mathrm{cos} \theta'_p \right)
	\nonumber\\& \times
	4\uppi (-\mathrm{i})^{n'}  \frac{-\mathrm{i}}{k_0}
	\times
	4\uppi (-\mathrm{i})^{n'}  
	\frac{k_0j_{n'}(k_sA)  	j'_{n'}(k_0A) -  k_s  j'_{n'}(k_sA)  	j_{n'}(k_0A)}{k_s^2 - k_0^2} A^2,
\end{align}

The expression we get here is without being averaged. It is possible to perform this averaging as we do in Subsection \ref{sec:Averaging the Green's function and the transition operator}, but obviously the expression will be complicated. The reason is that we are faced with four spherical harmonics and we must transform them into two spherical harmonics using Clebsch-Gordan coefficients if we want to apply orthogonality relation. Fortunately, when calculating the mean free paths, we can skip this procedure, which will be shown in Subsection \ref{sec:The transport mean free path}.

\section{The diffusion in small spheroidal particles group}
\label{sec:The diffusion in small spheroidal particles group}

\subsection{Transition operator in multiple scattering theory}
\label{sec:Transition operator in multiple scattering theory}

If we number all the particles in the scattering medium, we can get from Eq. (\ref{eq:G = G0 + G0 T G0}) that
\begin{align}
	\label{eq:G expansion in T}
	G  =  G_0  + \sum_{j=1}^{N}   G_0  T_j   G_0  
	+ \sum_{j=1}^{N} \sum_{k=1, k \ne j}^{N}
	G_0 T_k G_0  T_j   G_0  +  \cdots  ,
\end{align}
which is shown by Figure \ref{fig:multiple_scattering_in_spheroidal_particles_group}. $N$ is the total number of particles in the medium. $j$ and $k$ mean the $j$th and $k$th particles, respectively. We change the notation of the left hand side because it here represents the multiple scattering process, and in Eq. (\ref{eq:G = G0 + G0 T G0}) only one single particle is present. We have assumed that the light does not get into one particle twice, which corresponds to the so-called Twersky approximation \cite{Mishchenko2006}. Now we average Eq. (\ref{eq:G expansion in T}) and get
\begin{align}
	\label{eq:<G> first}
	\left\langle G \left( \mathbf{r}, \mathbf{r}_0 \right) \right\rangle = G_0 \left( \mathbf{r}, \mathbf{r}_0 \right) 
	+ \int \mathrm{d} \mathbf{r}_2 \mathrm{d} \mathbf{r}_1 G_0 \left( \mathbf{r}, \mathbf{r}_2 \right)   \Sigma \left( \mathbf{r}_2, \mathbf{r}_1 \right)  \left\langle G \left( \mathbf{r}_1, \mathbf{r}_0 \right) \right\rangle  ,
\end{align}
where
\begin{eqnarray}
	\label{eq:<Sigma> first}
	\Sigma   =  \sum_{j=1}^{N}    \left\langle   T_j   \right\rangle    + \sum_{j=1}^{N} \sum_{k=1, k \ne j}^{N}
	\left\langle  T_k G_0  T_j    \right\rangle  +  \cdots
\end{eqnarray}
is called the self-energy. Its second and higher terms represents those correlations between any two and more particles, and the first term means the uncorrelated part. When the particles are sparsely distributed in the medium, they hardly have impact on each other. Then we can naturally retain only the first term of $\Sigma$, which is called independent scattering. To make notations simpler, we use $\left\langle T \right\rangle $ to replace $\sum_{j=1}^{N}    \left\langle   T_j   \right\rangle$, which leads to
\begin{eqnarray}
	\label{eq:independent scattering}
	\Sigma   \simeq  \sum_{j=1}^{N}    \left\langle   T_j   \right\rangle   =  \left\langle   T   \right\rangle 
\end{eqnarray}
and
\begin{align}
	\label{eq:<G> second}
	\left\langle G \left( \mathbf{r}, \mathbf{r}_0 \right) \right\rangle &= G_0 \left( \mathbf{r}, \mathbf{r}_0 \right) 
	\nonumber\\
	+ \int &\mathrm{d} \mathbf{r}_2 \mathrm{d} \mathbf{r}_1 G_0 \left( \mathbf{r}, \mathbf{r}_2 \right)  \left\langle  T \left( \mathbf{r}_2, \mathbf{r}_1 \right) \right\rangle   \left\langle G \left( \mathbf{r}_1, \mathbf{r}_0 \right) \right\rangle  .
\end{align}
We call $\left\langle T \right\rangle $ the average transition operator. It describes the statistical result of one single transition. Next, we can take the Fourier transform of Eq. (\ref{eq:<G> second}) to get the Dyson equation
\begin{eqnarray}
	\label{eq:Dyson equation in <T>}
	\left\langle G \left( \mathbf{p} \right) \right\rangle = \frac{1}{p^2 - k_0^2 -    \Sigma \left( \mathbf{p} \right)    } = \frac{1}{p^2 - k_0^2 - \left\langle  T \left( \mathbf{p}, \mathbf{p} \right) \right\rangle  }.
\end{eqnarray}
Moreover, if the particles are monodisperse, namely identical, we get
\begin{align}
	\label{eq:<Tj> relation <T>}
	\left\langle T \left( \mathbf{r}, \mathbf{r}' \right)  \right\rangle  
	&=  	\frac{N}{V} \int \mathrm{d} \mathbf{r}_j  \int   
	\frac{\mathrm{d} \mathbf{p}}{(2\uppi)^3} \frac{\mathrm{d} \mathbf{p}'}{(2\uppi)^3} 
	\mathrm{e}^{\mathrm{i} \mathbf{p} \cdot (\mathbf{r} - \mathbf{r}_j) }
	T_j (\mathbf{p},  \mathbf{p}' ) 
	\mathrm{e}^{-\mathrm{i} \mathbf{p}' \cdot (\mathbf{r}' - \mathbf{r}_j) } 
	\nonumber\\ &
	= \frac{N}{V}  \int   
	\frac{\mathrm{d} \mathbf{p}}{(2\uppi)^3}  T_j (\mathbf{p},  \mathbf{p} )    \mathrm{e}^{\mathrm{i} \mathbf{p} \cdot (\mathbf{r} - \mathbf{r}') }      ,
\end{align}
where $N/V = n_j$ means that the number of particles are $N$ in volume $V$. This leads to 
\begin{align}
	\left\langle  T \left( \mathbf{p}, \mathbf{p}' \right)   \right\rangle  =  n_j
	T_j (\mathbf{p},  \mathbf{p} )  \delta  \left( \mathbf{p}, \mathbf{p}' \right).
\end{align}
Substituting Eq. (\ref{eq:final result of the transition operator}) into Eq. (\ref{eq:Dyson equation in <T>}) gives the average Green's function in the spheroidal particles group.

Now we consider averaging the product of retarded and advanced Green's functions $\left\langle GG^* \right\rangle $, which is called the intensity propagator. The propagator $\left\langle GG^* \right\rangle $ satisfies the Bethe-Salpeter equation
\begin{align}
	\label{eq:<GG*> first}
	\left\langle G \left( \mathbf{r}, \mathbf{r}_0 \right)  G^* \left( \mathbf{r}', \mathbf{r}'_0 \right) \right\rangle = & \left\langle G \left( \mathbf{r}, \mathbf{r}_0 \right) \right\rangle \left\langle  G \left( \mathbf{r}', \mathbf{r}'_0 \right) \right\rangle ^* 
	+ \int \mathrm{d} \mathbf{r}_2 \mathrm{d} \mathbf{r}_1 \mathrm{d} \mathbf{r}'_2 \mathrm{d} \mathbf{r}'_1  
	\nonumber\\  &\times
	\left\langle G \left( \mathbf{r}, \mathbf{r}_2 \right) \right\rangle \left\langle G \left( \mathbf{r}', \mathbf{r}'_2 \right) \right\rangle^*   
	U  ( \mathbf{r}_2, \mathbf{r}_1; \mathbf{r}'_2, \mathbf{r}'_1 )    
	\left\langle G \left( \mathbf{r}_1, \mathbf{r}_0 \right) G^* \left( \mathbf{r}'_1, \mathbf{r}'_0 \right) \right\rangle   .
\end{align}
$U$ is called the irreducible vertex, containing all the correlated scattering events. Just like the self-energy, $U$ also have infinite terms, and higher-order terms can also be neglected when the particles are sparse. This is called the ladder approximation, given by
\begin{eqnarray}
	\label{eq:U = <TT*>}
	U \simeq  \sum_{j=1}^{N}    \left\langle   T_j  T^*_j   \right\rangle   =  \left\langle   TT^*   \right\rangle .
\end{eqnarray}
Only one subscript $j$ appears in each angular bracket, which means that the retarded and advanced Green's function will have correlations only when they follow the same scattering sequence. Performing Fourier transformation of Eq. (\ref{eq:<GG*> first}) we get 
\begin{align}
	\label{eq:Phi the BSe}
	\Phi_{ \mathbf{p}, \mathbf{p}'} \left( \mathbf{q}  \right)  = (2\uppi)^3\left\langle G_{ \mathbf{p}+} \right\rangle  \left\langle G_{  \mathbf{p}-} \right\rangle^*  \delta_{\mathbf{p}, \mathbf{p}'}  
	+   \left\langle G_{ \mathbf{p}+} \right\rangle  \left\langle G_{  \mathbf{p}-} \right\rangle^*  \int \frac{ \mathrm{d} \mathbf{p}''}{(2\uppi)^3} U_{ \mathbf{p}, \mathbf{p}''} \left( \mathbf{q}  \right)  \Phi_{ \mathbf{p}'', \mathbf{p}'} \left( \mathbf{q}  \right), 
\end{align}
where
\begin{align}
	\label{eq:Phi_Fourier}
	\Phi_{ \mathbf{p}, \mathbf{p}'} \left( \mathbf{q}  \right)  \left( 2\uppi \right)^3  \delta \left( -\mathbf{p}_1 + \mathbf{p}_2 + \mathbf{p}_3 - \mathbf{p}_4 \right)   
	=   \left\langle G \left( \mathbf{p}_1, \mathbf{p}_2 \right)  G^* \left( \mathbf{p}_3, \mathbf{r}_4 \right) \right\rangle
\end{align}
and
\begin{align}
	\label{eq:U_Fourier}
	U_{ \mathbf{p}, \mathbf{p}'} \left( \mathbf{q}  \right)  \left( 2\uppi \right)^3  \delta \left( -\mathbf{p}_1 + \mathbf{p}_2 + \mathbf{p}_3 - \mathbf{p}_4 \right)   
	=   U  \left( \mathbf{p}_1, \mathbf{p}_2  ; \mathbf{p}_3, \mathbf{r}_4 \right)  
\end{align}
imply the momentum conservation, and we have made the change of variables $\mathbf{p}_1 = \mathbf{p} + \mathbf{q}/2$, $\mathbf{p}_2 = \mathbf{p}' + \mathbf{q}/2$, $\mathbf{p}_3 = \mathbf{p} - \mathbf{q}/2$. Also,
\begin{eqnarray}
	\label{eq:G+ G- definition}
	\left\langle G_{ \mathbf{p}+} \right\rangle = \left\langle G \left( \mathbf{p}+ \frac{\mathbf{q}}{2} \right)  \right\rangle,  \quad
	\left\langle G_{ \mathbf{p}-} \right\rangle =\left\langle G \left( \mathbf{p}- \frac{\mathbf{q}}{2} \right) \right\rangle.
\end{eqnarray}
Considering Eq. (\ref{eq:<Tj> relation <T>}), (\ref{eq:U = <TT*>}) and (\ref{eq:U_Fourier}), we get
\begin{align}
	\label{eq:U = <TT*> in Fourier}
	U_{\mathbf{p}, \mathbf{p}'} \left( \mathbf{q} \right)  \simeq   n_j     \left\langle   T_j \left( \mathbf{p}+ , \mathbf{p}'+  \right)  T^*_j \left( \mathbf{p}-  , \mathbf{p}'-   \right)    \right\rangle     .
\end{align}

\subsection{The diffusion approximation and the far-field approximation}
\label{sec:far field approximation}

The diffusion approximation is
\begin{align}
	\label{eq:diffusion approximation}
	q \to 0
\end{align}
in Eq. (\ref{eq:Phi the BSe}). Because $q$ is the conjugate variable of propagation distance, this approximation means that the wave propagates for a much longer distance than the scale of dielectric inhomogeneities. In this case we can set $q = 0$ in $\left\langle G_{\mathbf{p}+} \right\rangle $, $\left\langle G_{\mathbf{p}-} \right\rangle $ and $U_{\mathbf{p}, \mathbf{p}'} \left( \mathbf{q} \right)$.

Moreover, the free space Green's function in the far-field region has the asymptotic form that
\begin{align}
	G_0 \left( \mathbf{r} - \mathbf{r}' \right) = 
	\frac{\mathrm{exp}\left( \mathrm{i}k_0 \left| \mathbf{r} - \mathbf{r}' \right|  \right) }{4\uppi \left| \mathbf{r} - \mathbf{r}' \right|}
	\simeq   \frac{\mathrm{exp}\left( \mathrm{i}k_0 r  \right) }{4\uppi r}   
	\mathrm{exp}\left( -\mathrm{i}k_0  \hat{\mathbf{r}} \cdot \mathbf{r}'   \right),
\end{align}
whose term $\mathrm{exp}\left( -\mathrm{i}k_0  \hat{\mathbf{r}} \cdot \mathbf{r}'   \right)$ could be considered as a Fourier transformation kernel.
By using Eq. (\ref{eq:G = G0 + G0 T G0}) and (\ref{eq:Tj transformation definition}) we get
\begin{align}
	\label{eq:far field approximation}
	G_s\left( \mathbf{r}, \mathbf{r}'  \right) =  
	\frac{\mathrm{exp}\left( \mathrm{i}k_0 r  \right) }{4\uppi r}  T_j \left( k_0\hat{\mathbf{r}}, k_0\hat{\mathbf{r}'} \right) \frac{\mathrm{exp}\left( \mathrm{i}k_0 r'  \right) }{4\uppi r'},
\end{align}
where $G_s = G_e - G_0$. Eq. (\ref{eq:far field approximation}) has shown that $p = p' = k_0$ holds in $T_j \left( \mathbf{p}, \mathbf{p}' \right) $ for far-field region. This reveals that the particles are sparse and each of them is in the far-field regions of any other particles. In fact, there is another approximation called the on-shell approximation that also leads to $p = p' = k_0$ \cite{Piraud_2013}. It tells us that, because 
\begin{align}
	\left\langle G_{ \mathbf{p}+} \right\rangle  \left\langle G_{  \mathbf{p}-} \right\rangle^* =  \frac{\left\langle G_{ \mathbf{p}+} \right\rangle - \left\langle G_{ \mathbf{p}-} \right\rangle} {\left\langle G^{-1}_{ \mathbf{p}-} \right\rangle - \left\langle G^{-1}_{ \mathbf{p}+} \right\rangle}
	=  \frac{ \Delta G _{ \mathbf{p}} (\mathbf{q})  } { \left(  \Sigma_{ \mathbf{p}+} - \Sigma_{ \mathbf{p}-} \right)  - 2\mathbf{p} \cdot \mathbf{q}  }  
\end{align} 
and Eq. (\ref{eq:diffusion approximation}), the Bethe-Salpeter Eq. (\ref{eq:Phi the BSe}) is limited to $\Delta G _{ \mathbf{p}} (\mathbf{0})  =  -\uppi \delta (k_0^2 - p^2)$. Therefore, the on-shell approximation and the far-field approximation both result in
\begin{align}
	\label{eq:U = <TT*> final}
	U_{\mathbf{p}, \mathbf{p}'} \left( \mathbf{q} \right)  \approx  
	U_{k_0\hat{\mathbf{p}}, \, k_0\hat{\mathbf{p}'}} \left( \mathbf{0} \right)
	=  n_j    \left|    T_j \left( k_0\hat{\mathbf{p}}, k_0\hat{\mathbf{p}'}  \right)  \right| ^2    .
\end{align}
This is why we calculate Eq. (\ref{eq:final result of the transition operator}) by setting $p = p' = k_0$.

\subsection{The transport mean free path}
\label{sec:The transport mean free path}

The approaches to solving the Bethe-Salpeter equation in the diffusion approximation are widely discussed \cite{Akkermans2007,Carminati2021,Rammer2004,Tiggelen_2016,LAGENDIJK1996143}. We will not repeat here but directly give the solution of Eq. (\ref{eq:Phi the BSe}), which is
\begin{align}
	\label{eq:Quantum probability}
	\Phi  (\mathbf{q})  =  \frac{1}{D^* q^2}.
\end{align}
$D^* = v\ell^*/3$ is the diffusion constant, with the transport mean free path $\ell^*$ given by
\begin{align}
	\ell^* = \frac{\ell}{1-g}.
\end{align}
The scattering mean free path $\ell$ and the anisotropy factor $g$ is given by \cite{Carminati2021}
\begin{align}
	\label{eq:ell the definition}
	\frac{1}{\ell} = \frac{n_j}{16\uppi^2}  \int \left|    T_j \left( k_0\hat{\mathbf{p}}, k_0\hat{\mathbf{p}'}  \right)  \right| ^2  \mathrm{d} \hat{\mathbf{p}},
\end{align}
\begin{align}
	\label{eq:<cos> the definition}
	g = \frac{n_j\ell}{16\uppi^2}  \int  \left(   \hat{\mathbf{p}} \cdot \hat{\mathbf{p}'} \right)  \left|    T_j \left( k_0\hat{\mathbf{p}}, k_0\hat{\mathbf{p}'}  \right)  \right| ^2  \mathrm{d} \hat{\mathbf{p}}.
\end{align}
The mean free paths $\ell$ and $\ell^*$ are very important quantities, because they connect the single scattering process to the multiple scattering process.

Now let us consider how to calculate $\ell^{-1}$ and $g$. In the derivation of $\left\langle T_j \right\rangle $ (Subsection \ref{sec:Averaging the Green's function and the transition operator}), we use Wigner D function to represent the rotation of the coordinate system, and perform averaging over Euler angles by using the orthogonality of Wigner D function. When averaging over Euler angles, we keep the angle $\hat{\mathbf{p}} - \hat{\mathbf{p}}'$ unchanged \footnote{In fact, we do this in real space. Considering the Fourier kernel in Eq. (\ref{eq:Stratton Fourier of spherical harmonics}) is $\mathrm{exp}( - \mathrm{i}\mathbf{p} \cdot \mathbf{r})$, and this kernel is unaffected by the rotation, we deduce that the angles between $(\theta, \phi)$ and $(\theta_p, \phi_p)$ is invariant with the rotation, and the averaging in reciprocal space has the same form with the averaging in real space.},    and retrieve all values of $(\hat{\mathbf{p}}+\hat{\mathbf{p}}')/2$. For now, if we want to calculate the mean free paths, we need to retrieve all values of the angle $\hat{\mathbf{p}} - \hat{\mathbf{p}}'$. As a matter of fact, we retrieve all values of $\hat{\mathbf{p}} $ and $ \hat{\mathbf{p}}'$ in these two steps. This means that we can skip the averaging and directly calculate the integral over $\hat{\mathbf{p}}$ and $\hat{\mathbf{p}}'$, which makes the calculation much simpler. So we get
\begin{align}
	\frac{1}{\ell} &= \frac{n_j}{16\uppi^2}  \int  \left\lbrace \frac{1}{4\uppi} \int \left|    T_j \left( k_0\hat{\mathbf{p}}, k_0\hat{\mathbf{p}'}  \right)  \right| ^2 \mathrm{d}  \hat{\mathbf{p}'} \right\rbrace  \mathrm{d} \hat{\mathbf{p}}
	\nonumber\\
	&= \frac{n_j}{64\uppi^3}  \int \left|    T_j \left( k_0\hat{\mathbf{p}}, k_0\hat{\mathbf{p}'}  \right)  \right| ^2 \mathrm{d}  \hat{\mathbf{p}'}  \mathrm{d}  \hat{\mathbf{p}}.
\end{align}
Substituting the expression of $ | T_j |^2  $, namely Eq. (\ref{eq:Tj*Tj the Fourier}), we get 
\begin{align}
	\label{eq:ell the final result}
	\frac{1}{\ell} &= 4\uppi n_j  
	k_0^2 (\delta \epsilon_p)^2    k_s^2 
	\sum_{n=0}^{\infty}        
	\sum_{n'=0}^{\infty} \sum_{m=0}^{ \mathrm{min} \{n,n'\} } \left( 2-\delta_{m,0}  \right) ^2   C_{mn} C^*_{mn'}  N^2_{mn}(0)  N^2_{mn'}(0)
	\nonumber\\& \times
	\frac{k_0j_n(k_sA)  	j'_n(k_0A) -  k_s  j'_n(k_sA)  	j_n(k_0A)}{k_s^2 - k_0^2} A^2 
	\frac{k_0j_{n'}(k_sA)  	j'_{n'}(k_0A) -  k_s  j'_{n'}(k_sA)  	j_{n'}(k_0A)}{k_s^2 - k_0^2} A^2
	\nonumber\\ & \times
	\sum_{l_1=m}^{\infty}   \sum_{l_2=m}^{\infty}
	\frac{d_{l_1-m}^{mn} (k_0a)}{N_{ml_1}(k_0a)}
	\frac{d_{l_2-m}^{mn} (k_sa)}{N_{ml_2}(k_sa)}
	\frac{d_{l_1-m}^{mn'} (k_0a)}{N_{ml_1}(k_0a)}
	\frac{d_{l_2-m}^{mn'} (k_sa)}{N_{ml_2}(k_sa)}.
\end{align}
where we have used Eq. (\ref{eq:orthogonality over two angles}).
And the result of anisotropy factor is
\begin{align}
	\label{eq:g the final result}
	g &= \frac{n_j\ell}{64\uppi^3}  \int       \left[  \mathrm{cos} \theta_p  \mathrm{cos} \theta'_p  +  \mathrm{sin} \theta_p  \mathrm{sin} \theta'_p \mathrm{cos} \left( \phi_p - \phi'_p  \right)  \right]  \left|    T_j \left( k_0\hat{\mathbf{p}}, k_0\hat{\mathbf{p}'}  \right)  \right| ^2 \mathrm{d} \hat{\mathbf{p}} \mathrm{d} \hat{\mathbf{p}'}   
	\nonumber\\ &
	= 4\uppi  n_j\ell   
	k_0^2 (\delta \epsilon_p)^2     k_s^2
	  \sum_{n=0}^{\infty} 
	   \sum_{n'=0}^{\infty}  
	   \sum_{m=0}^{n} \left( 2-\delta_{m,0}  \right)     C_{mn}   N^2_{mn}(0)
	   \sum_{m'=0}^{n'}
	    \left( 2-\delta_{m',0}  \right)     C^*_{m'n'}  N^2_{m'n'}(0)
	\nonumber\\& \quad \times   
	\frac{k_0 j_n(k_sA)  	j'_n(k_0A) -  k_s  j'_n(k_sA)  	j_n(k_0A)}{k_s^2 - k_0^2} A^2 
	\frac{k_0j_{n'}(k_sA)  	j'_{n'}(k_0A) -  k_s  j'_{n'}(k_sA)  	j_{n'}(k_0A)}{k_s^2 - k_0^2} A^2
	\nonumber\\ &  \quad\times
	\sum_{l_1=m}^{\infty}  \mathrm{i}^{l_1 } 
	\frac{ d_{l_1-m}^{mn} (k_0a) }{N_{ml_1}(k_0a)}
	\sum_{l_2=m}^{\infty}  \mathrm{i}^{l_2 } 
	\frac{ d_{l_2-m}^{mn} (k_sa) }{N_{ml_2}(k_sa)}
	\sum_{l_3=m'}^{\infty}  (-\mathrm{i})^{l_3 } 
	\frac{ d_{l_3-m'}^{m'n'} (k_0a) }{N_{m'l_3}(k_0a)}
	\sum_{l_4=m'}^{\infty}  (-\mathrm{i})^{l_4 } 
	\frac{ d_{l_4-m'}^{m'n'} (k_sa)  }{N_{m'l_4}(k_sa)}	
	\nonumber\\  &  \quad\times
	  \left[ \delta_{m'm}   f^{(=)}_{l_1,l_3}  
	f^{(=)}_{l_2,l_4}    + \frac{ 1 }{2}    \left( \delta_{m',m+1} f^{(+)}_{l_1,l_3} f^{(+)}_{l_2,l_4} +  \delta_{m',m-1} f^{(-)}_{l_1,l_3}  f^{(-)}_{l_2,l_4}    \right)       \right]   ,
\end{align}
where we have used Eq. (\ref{eq:intersection angle}) in the first equal-sign. In the second equal-sign, we first perform the integration over $(\phi_p+\phi'_p)/2$ and $\phi_p-\phi'_p$, and then perform the integration over $\theta_p$ and $\theta'_p$. The following formulae have also been used.
\begin{align}
	(2l+1)\sqrt{1-\mathrm{cos}^2\theta }  P_l^m  =  P_{l-1}^{m+1}  -   P_{l+1}^{m+1} ,
\end{align}
\begin{align}
	\int_{-1}^{1}   P_l^m  P_{l'}^m  \mathrm{d} x =  N^2_{ml} (0) \delta_{l' l}  ,
\end{align}
\begin{align}
	f^{(+)}_{l_2,l_4}  &= \int_{-1}^{1}   \mathrm{sin} \theta'_p   
	\overline{P}_{l_2}^m\left( \mathrm{cos} \theta'_p \right)
	\overline{P}_{l_4}^{m+1}\left( \mathrm{cos} \theta'_p \right)
	\mathrm{d} \mathrm{cos} \theta'_p 
	\nonumber\\&
	= \frac{1}{N_{ml_2}(0) N_{m+1, l_4}(0) }  \int_{0}^{\uppi}   \left(  1-\mathrm{cos}^2 \theta'_p  \right) 
	\frac{P_{l_2-1}^{m+1}\left( \mathrm{cos} \theta'_p \right) - P_{l_2+1}^{m+1}\left( \mathrm{cos} \theta'_p \right)}{(2l_2+1)\sqrt{1-\mathrm{cos}^2\theta }}
	P_{l_4}^{m+1}\left( \mathrm{cos} \theta'_p \right)
	\mathrm{d}  \theta'_p
	\nonumber\\&
	= \frac{1}{N_{ml_2}(0) N_{m+1, l_4}(0) }  \int_{-1}^{1}  
	\frac{P_{l_2-1}^{m+1}\left( \mathrm{cos} \theta'_p \right) - P_{l_2+1}^{m+1}\left( \mathrm{cos} \theta'_p \right)}{(2l_2+1) }
	P_{l_4}^{m+1}\left( \mathrm{cos} \theta'_p \right)
	\mathrm{d} \mathrm{cos}  \theta'_p
	\nonumber\\&
	= \frac{N^2_{m+1,l_4} (0) \delta_{l_2-1, l_4} - N^2_{m+1,l_4} (0) \delta_{l_2+1, l_4} }{ (2l_2+1) N_{ml_2}(0) N_{m+1, l_4}(0) }   ,
\end{align}
\begin{align}
	f^{(-)}_{l_2,l_4}  &= \int_{-1}^{1}   \mathrm{sin} \theta'_p   
	\overline{P}_{l_2}^m\left( \mathrm{cos} \theta'_p \right)
	\overline{P}_{l_4}^{m-1}\left( \mathrm{cos} \theta'_p \right)
	\mathrm{d} \mathrm{cos} \theta'_p 
	\nonumber\\&
	= \frac{1}{N_{ml_2}(0) N_{m-1, l_4}(0) }  \int_{0}^{\uppi}   \left(  1-\mathrm{cos}^2 \theta'_p  \right) P_{l_2}^{m}\left( \mathrm{cos} \theta'_p \right)
	\frac{P_{l_4-1}^{m}\left( \mathrm{cos} \theta'_p \right) - P_{l_4+1}^{m}\left( \mathrm{cos} \theta'_p \right)}{(2l_4+1)\sqrt{1-\mathrm{cos}^2\theta }}
	\mathrm{d}  \theta'_p
	\nonumber\\&
	= \frac{1}{N_{ml_2}(0) N_{m-1, l_4}(0) }  \int_{-1}^{1}  
	P_{l_2}^{m}\left( \mathrm{cos} \theta'_p \right)
	\frac{P_{l_4-1}^{m}\left( \mathrm{cos} \theta'_p \right) - P_{l_4+1}^{m}\left( \mathrm{cos} \theta'_p \right)}{(2l_4+1)\sqrt{1-\mathrm{cos}^2\theta }}
	\mathrm{d} \mathrm{cos}  \theta'_p
	\nonumber\\&
	= \frac{ N^2_{ml_2} (0) \delta_{l_2, l_4-1}  -   N^2_{ml_2} (0) \delta_{l_2, l_4+1}   }{(2l_4+1)  N_{ml_2}(0) N_{m-1, l_4}(0) }   .
\end{align}
\begin{align}
	(2l+1) x P_l^m = (l+m) P_{l-1}^m   +   (l-m+1)P_{l+1}^m   ,
\end{align}
\begin{align}
	f^{(=)}_{l_2,l_4}  &= \int_{-1}^{1}   \mathrm{cos} \theta'_p   
	\overline{P}_{l_2}^m\left( \mathrm{cos} \theta'_p \right)
	\overline{P}_{l_4}^{m}\left( \mathrm{cos} \theta'_p \right)
	\mathrm{d} \mathrm{cos} \theta'_p 
	\nonumber\\&
	= \frac{1}{N_{ml_2}(0) N_{m-1, l_4}(0) }    \int_{-1}^{1}   x
	 P_{l_2}^m\left( x \right)
	P_{l_4}^{m}\left( x \right)
	\mathrm{d} x
	\nonumber\\&
	= \frac{1}{  (2l_2+1) N_{ml_2}(0) N_{m-1, l_4}(0) }    \int_{-1}^{1}   
	\left[ (l_2+m)P_{l_2-1}^m (x) + (l_2-m+1)P_{l_2+1}^m (x) \right] 
	P_{l_4}^{m}\left( x \right)
	\mathrm{d} x
	\nonumber\\&
	= \frac{ (l_2+m) N^2_{m,l_4} \delta_{l_2-1, l_4}  + (l_2-m+1) N^2_{m,l_4}  \delta_{l_2+1, l_4} }{  (2l_2+1) N_{ml_2}(0) N_{m-1, l_4}(0) }    .
\end{align}

Finally, Eq. (\ref{eq:ell the final result}) and (\ref{eq:g the final result}) have given the expression of scattering mean free path and the anisotropy factor, respectively, for a group of monodisperse spheroidal particles.

\section{Conclusion}
\label{sec:conclusion}

Multiple scattering of coherent wave in non-spherical particles is an important issue because most small particles are not spherical in natural environment like seawater and atmosphere. In this manuscript we have used the expansion of scattering Green's function to derive the mean free paths concerning multiple scattering. First, we show how to get the average transition operator from the scattering Green's function for a group of dielectric spheroids with random orientations. The Green's function is expanded in a prolate spheroidal coordinate system, and then transformed into expansions of spherical wavefunctions. The averaging of a transition operator over Euler angles has been performed with the orthogonality of Wigner D function. In the second part, we have demonstrated the diffusion of scalar coherent waves in a group of spheroidal particles. We apply the diffusion approximation and far-field approximation to the irreducible vertex of the Bethe-Salpeter equation, to show the relation between the mean free paths and the transition operator. Having noticed the importance of the product of a transition operator and its conjugate, we show an alternative way to calculate the mean free paths from a transition operator without the averaging.

\newpage

\begin{appendices}

\section{Appendix: Spheroidal functions}
\label{sec:spheroidal functions}

The spheroidal radial function $R_{mn}^{(i)}  \left( k_0 a, \mathrm{cosh} \xi \right) $ satisfies \cite{LiLewei2001}
\begin{align}
	\label{eq:R differential equation}
	\displaystyle\frac{\mathrm{d}^2 R_{mn}^{(i)}  }{\mathrm{d}\xi^2}  +\mathrm{coth} \xi \frac{\mathrm{d} R_{mn}^{(i)}  }{\mathrm{d}\xi}   
	+\left( k_0^2 a^2 \mathrm{sinh}^2 \xi - n(n+1) - \frac{m^2}{\mathrm{sinh}^2\xi} \right) R_{mn}^{(i)}  = 0,
\end{align}
and the spheroidal angular function $\mathscr{P}_n^m  \left( k_0 a, \mathrm{cos}\vartheta \right) $ satisfies
\begin{align}
	\label{eq:P differential equation}
	\displaystyle\frac{\mathrm{d}^2 \mathscr{P}_n^m }{\mathrm{d}\vartheta^2}  +\mathrm{cot} \vartheta \frac{\mathrm{d} \mathscr{P}_n^m  }{\mathrm{d}\vartheta}   
	+\left( k_0^2 a^2 \mathrm{sin}^2 \vartheta + n(n+1) - \frac{m^2}{\mathrm{sin}^2\vartheta} \right) \mathscr{P}_n^m  = 0 .
\end{align}
The normalized spheroidal angular function $\overline{\mathscr{P}}_n^m$ is given by
\begin{align}
	\overline{\mathscr{P}}_n^m  \left( k_0 a, \mathrm{cos}\vartheta \right)  =   N^{-1}_{mn}  (k_0 a)    \mathscr{P}_n^m  \left( k_0 a, \mathrm{cos}\vartheta \right),
\end{align}
where the normalization factor is given by
\begin{eqnarray}
	\label{eq:Nmn}
	N_{mn} \left(k_0a \right)  = 2 {\sum_{l=0,1}^{\infty} }' \frac{\left( l+2m \right)! }{(2l+2m+1)!} \left[ d_l^{mn} \left(k_0a \right) \right]^2.
\end{eqnarray}
Specially, \cite{Farafonov_2016}
\begin{align}
	\label{eq:Nmn(0)}
	N_{mn} \left( 0 \right)  =  \sqrt{\frac{2}{2n+1} \frac{(n+m)!}{(n-m)!} }
\end{align}
is the normalization factor of the associated Legendre function.
In Eq. (\ref{eq:Nmn}), the prime over the summation sign indicates that the summation is over only even values of $l$ when $n-m$ is even, and over only odd values of $l$ when $n-m$ is odd. The coefficient $d_l^{mn}$ is determined by the following recursion formulae \cite{adelman2014softwarecomputingspheroidalwave}
\begin{align}
	\alpha_l    d_{l+2}^{mn} (k_0a)  +\left( \beta_l - \lambda_{mn} (k_0a) \right)  d_l^{mn} (k_0a)  +   \gamma_l    d_{l-2}^{mn} (k_0a)   =  0  ,
\end{align}
where
\begin{align}
	\label{eq:alpha_l}
	\alpha_l = \frac{(2m+l+2)(2m+l+1)}{(2m+2l+3)(2m+2l+5)}  (k_0a)^2,
\end{align}
\begin{align}
	\label{eq:beta_l}
	\beta_l = (m+l)(m+l+1)  + \frac{2(m+l)(m+l+1) - 2m^2 - 1}{(2m+2l-1)(2m+2l+3)}  (k_0a)^2,
\end{align}
\begin{align}
	\label{eq:gamma_l}
	\gamma_l  = \frac{l(l-1)}{(2m+2l-3)(2m+2l-1)}  (k_0a)^2.
\end{align}

\section{Appendix: Coefficients $A_{mn}$ and $C_{mn}$}
\label{sec:Amn and Cmn}

The boundary condition of scalar Green's function for a dielectric spheroid could be expressed by \cite{Carminati2021,Tai_1971}
\begin{align}
	\label{eq:boundary condition}
	G_e^{(22)} \left( \mathbf{r}, \mathbf{r}' \right)  &= G_e^{(12)} \left( \mathbf{r}, \mathbf{r}' \right),
	\nonumber\\
	\nabla G_e^{(22)} \left( \mathbf{r}, \mathbf{r}' \right) \cdot \hat{\mathbf{n}} &= 	\nabla G_e^{(12)} \left( \mathbf{r}, \mathbf{r}' \right) \cdot \hat{\mathbf{n}},
\end{align}
when $\xi = \xi_0$. This boundary condition means that the limiting values of the scalar Green's function as $\mathbf{r}$ approaches the interface from region 1 and from region 2 is equal. Applying Eq. (\ref{eq:boundary condition}) to Eq. (\ref{eq:composition of Ge first}) yields
\begin{align}
	\label{eq:An Cn relation 1}
	R_{mn}^{(3)} \left( k_s a, \mathrm{cosh} \xi_0  \right) + 
	A_{mn}  R_{mn}^{(1)} \left( k_s a, \mathrm{cosh} \xi_0  \right)
	= C_{mn} R_{mn}^{(3)} \left( k_0 a, \mathrm{cosh} \xi_0  \right),
\end{align}
\begin{align}
	\label{eq:An Cn relation 2}
	\frac{R_{mn}^{(3)'} \left( k_s a, \mathrm{cosh} \xi_0  \right) }{a \sqrt{\mathrm{sinh}^2 \xi_0 + \mathrm{sin}^2 \vartheta }   }   + 
	A_{mn}   \frac{  R_{mn}^{(1)'} \left( k_s a, \mathrm{cosh} \xi_0  \right)  }{a \sqrt{\mathrm{sinh}^2 \xi_0 + \mathrm{sin}^2 \vartheta }  }
	= C_{mn}    \frac{  R_{mn}^{(3)'} \left( k_0 a, \mathrm{cosh} \xi_0  \right) }{a \sqrt{\mathrm{sinh}^2 \xi_0 + \mathrm{sin}^2 \vartheta }  } ,
\end{align}
where the prime indicates derivative with respect to $\xi$. The denominators in Eq. (\ref{eq:An Cn relation 2}) cancel because they have the same $\vartheta$. Thus, we get the coefficients
\begin{eqnarray}
	\label{eq:An coefficients}
	A_{mn} =  \frac{R_{mn}^{(3)'} \left( k_s a  \right)_{\xi_0 } R_{mn}^{(3)} \left( k_0 a  \right)_{\xi_0 } - R_{mn}^{(3)'} \left( k_0 a  \right)_{\xi_0 } R_{mn}^{(3)} \left( k_s a  \right)_{\xi_0 } }
	{R_{mn}^{(3)'} \left( k_0 a  \right)_{\xi_0 } R_{mn}^{(1)} \left( k_s a  \right)_{\xi_0 } - R_{mn}^{(1)'} \left( k_s a  \right)_{\xi_0 } R_{mn}^{(3)} \left( k_0 a  \right)_{\xi_0 }  },
\end{eqnarray}
\begin{eqnarray}
	\label{eq:Cn coefficients}
	C_{mn}  =  \frac{R_{mn}^{(3)'} \left( k_0 a  \right)_{\xi_0 } R_{mn}^{(3)} \left( k_s a  \right)_{\xi_0 } - R_{mn}^{(3)'} \left( k_s a  \right)_{\xi_0 } R_{mn}^{(3)} \left( k_0 a  \right)_{\xi_0 }  }
	{R_{mn}^{(1)'} \left( k_s a  \right)_{\xi_0 } R_{mn}^{(3)} \left( k_s a  \right)_{\xi_0 } - R_{mn}^{(3)'} \left( k_s a  \right)_{\xi_0 } R_{mn}^{(1)} \left( k_s a  \right)_{\xi_0 }  },
\end{eqnarray}
where the radial function $R_{mn}^{(i)} \left( k_0 a \right)_{\xi_0 } = R_{mn}^{(i)} \left( k_0 a, \mathrm{cosh} \xi_0  \right)$.

\section{Appendix: Fourier transformation of the scalar spherical wave functions}

Stratton has given the formula (see Page 410 of Ref. \cite{Stratton1941})
\begin{align}
	\label{eq:Stratton Fourier of spherical harmonics}
	j_n(pr) P_n^m (\mathrm{cos}\theta_p)  \mathrm{e}^{\mathrm{i} m \phi_p} = \frac{\mathrm{i}^{-n}}{4\uppi}  \int_{0}^{\uppi}  (\mathrm{sin}\theta) \mathrm{d}  \theta  
	\int_{-\uppi}^{\uppi} \mathrm{d}  \phi     \left[  \mathrm{e}^{-\mathrm{i} pr\mathrm{cos}\Theta }  P_n^m (\mathrm{cos}\theta)  \mathrm{e}^{\mathrm{i} m \phi} \right] ,
\end{align}
where 
\begin{eqnarray}
	\mathrm{cos} \Theta  = \mathrm{cos} \theta_p  \mathrm{cos} \theta  +  \mathrm{sin} \theta_p  \mathrm{sin} \theta \mathrm{cos} \left( \phi_p - \phi  \right) .
\end{eqnarray}
Eq. (\ref{eq:Stratton Fourier of spherical harmonics}) shows the Fourier transformation of a spherical harmonics by setting $\Theta$ as the angle between the real-space vector $\mathbf{r} = (r,\theta,\phi)$ and the Fourier-space vector $\mathbf{p} = (p,\theta_p,\phi_p)$. Moreover, using the following integral formulae \cite{Table_of_integrals, WangZhuxi}
\begin{align}
	\int r^2 z^{(i_1)}_n(kr)  z^{(i_2)}_n(pr)  \mathrm{d}r   
	= \frac{pz^{(i_1)}_n(kr)  z^{(i_2)'}_n(pr)  -  kz^{(i_1)'}_n(kr)  z^{(i_2)}_n(pr)}{k^2 - p^2}  r^2 ,
\end{align}
\begin{align}
	\int_{0}^{\infty} r^2 z^{(i_1)}_n(kr)  z^{(i_2)}_n(pr)  \mathrm{d}r
	= \frac{\uppi\delta (k-p)}{2k^2},
\end{align}
we get
\begin{align}
	\label{eq:appendix formula 1}
	\int_{0}^{\infty}  r^2 \mathrm{d}r  \int_{-1}^{1} &  \mathrm{d}  \mathrm{cos}\theta   \int_{\uppi}^{\uppi}  \mathrm{d}  \phi     \left[  \mathrm{e}^{\mathrm{i} \mathbf{p}\cdot \mathbf{r} }  z_n(k_0r)  P_n^m (\mathrm{cos}\theta)  \mathrm{e}^{\mathrm{i} m \phi} \right]  
	\nonumber\\ = &
	4\uppi \mathrm{i}^{n} P_n^m (\mathrm{cos}\theta_p)  \mathrm{e}^{\mathrm{i} m \phi_p} \int_{0}^{\infty}  r^2 z_n(k_0r)  	j_n(pr)  \mathrm{d}r
	\nonumber\\[2mm] = &
	4\uppi \mathrm{i}^{n} P_n^m (\mathrm{cos}\theta_p)  \mathrm{e}^{\mathrm{i} m \phi_p} \frac{\uppi}{2k_0^2} \delta (k_0-p)  ,
\end{align}
\begin{align}
	\label{eq:appendix formula 2}
	\int_{0}^{a}  r^2 \mathrm{d}r  \int_{-1}^{1}  & \mathrm{d}  \mathrm{cos}\theta   \int_{\uppi}^{\uppi}  \mathrm{d}  \phi     \left[  \mathrm{e}^{\mathrm{i} \mathbf{p}\cdot \mathbf{r} }  z_n(k_0r)  P_n^m (\mathrm{cos}\theta)  \mathrm{e}^{\mathrm{i} m \phi} \right]  
	\nonumber\\  &=
	4\uppi \mathrm{i}^{n} P_n^m (\mathrm{cos}\theta_p)  \mathrm{e}^{\mathrm{i} m \phi_p} \int_{0}^{a}  r^2 z_n(k_0r)  	j_n(pr)  \mathrm{d}r
	\nonumber\\[2mm] &= 
	4\uppi \mathrm{i}^{n} P_n^m (\mathrm{cos}\theta_p)  \mathrm{e}^{\mathrm{i} m \phi_p} 
	\frac{pz_n(k_0a)  	j'_n(pa) -  k_0  z'_n(k_0a)  	j_n(pa)}{k_0^2 - p^2} a^2   .
\end{align}

\section{Appendix: Calculation of $\left\langle { D_{m'k'}^{l'} }^*  D_{mk}^{l}  \right\rangle _{\alpha,\beta,\gamma} $}

In this appendix we show the calculation of $\left\langle { D_{m'k'}^{l'} }^*   D_{mk}^{l}  \right\rangle _{\alpha,\beta,\gamma} $ (for more details please see Ref. \cite{Mishchenko2002}).
For any function $f\left( \alpha, \beta, \gamma \right) $, its average over Euler angles is expressed by
\begin{align}
	\label{eq:average alpha beta gamma}
	\left\langle f\left( \alpha, \beta, \gamma \right) \right\rangle _{\alpha, \beta, \gamma}    = 
	\int_{0}^{2\uppi} \mathrm{d}  \alpha \int_{0}^{\uppi}  \mathrm{sin} \beta \mathrm{d} \beta \int_{0}^{2\uppi}  \mathrm{d} \gamma 
	f\left( \alpha, \beta, \gamma \right)  
	p\left( \alpha, \beta, \gamma \right).
\end{align}
$p\left( \alpha, \beta, \gamma \right)$ is the probability density function of $\alpha, \beta, \gamma$. We assume the orientation of each particle is totally random, and namely the Euler angles are uniformly distributed, which means that 
\begin{eqnarray}
	p \left( \alpha, \beta, \gamma \right) = \frac{1}{8\uppi^2} . 
\end{eqnarray}
The definition of Wigner D function
\begin{eqnarray}
	\label{eq:Wigner D function}
	D_{mk}^{l} \left( \alpha, \beta, \gamma \right) = \mathrm{e}^{-\mathrm{i} m \alpha } d^l_{mk} (\beta)  \mathrm{e}^{-\mathrm{i} k \gamma  }
\end{eqnarray}
where the Wigner d function satisfies 
\begin{eqnarray}
	\label{eq:Wigner d function orthogonality}
	\int_{0}^{\uppi} \mathrm{sin} \beta \mathrm{d} \beta d^{l}_{mk} (\beta) d^{l'}_{m'k'} (\beta)  =  \frac{2}{2l+1} \delta_{l'l}.
\end{eqnarray}
Considering 
\begin{eqnarray}
	\label{eq:orthogonality over two angles}
	\int_{0}^{2\uppi}  \mathrm{d}  \alpha  \mathrm{exp}  \left[ -\mathrm{i} \left( m - m' \right) \alpha \right] = 2\uppi \delta_{m'm} 
\end{eqnarray}
and Eq. (\ref{eq:average alpha beta gamma}), (\ref{eq:Wigner D function}), and (\ref{eq:Wigner d function orthogonality}), we get
\begin{align}
	\label{eq:<D*D> calculation}
	&\left\langle { D_{m'k'}^{l'} }^* \left( \alpha, \beta, \gamma \right) D_{mk}^{l} \left( \alpha, \beta, \gamma \right)  \right\rangle  _{\alpha,\beta,\gamma}   
	\nonumber\\&  \qquad\qquad
	=   \frac{1}{8\uppi^2}
	\int_{0}^{2\uppi}   \mathrm{d} \alpha   
	\mathrm{e}^{\mathrm{i} m' \alpha } \mathrm{e}^{-\mathrm{i} m \alpha } 
	\int_{0}^{2\uppi}  \mathrm{d}  \gamma
	\mathrm{e}^{\mathrm{i} k' \gamma  }  \mathrm{e}^{-\mathrm{i} k \gamma  }
	\int_{0}^{\uppi}   \mathrm{sin} \beta \mathrm{d} \beta 
	d^{l}_{mk} (\beta) d^{l'}_{m'k'} (\beta)  
	\nonumber\\ &  \qquad\qquad  =
	\frac{\delta_{m'm} \delta_{k'k} \delta_{l' l}}{2l+1}  .
\end{align}

\section{Appendix: The expansion of $G_0$ }
\label{sec:spherical}

Eq. (\ref{eq:G0 expansion}) reads
\begin{align}
	\label{eq:G0 appendix expansion 1}
	G_0 \left( \mathbf{r}, \mathbf{r}' \right) 
	&= \frac{\mathrm{i}k_0}{2\uppi} \sum_{m,n}  \left( 2-\delta_{m,0} \right)   
	\overline{\mathscr{P}}_n^m  \left( k_0 a, \mathrm{cos} \vartheta  \right) 
	\overline{\mathscr{P}}_n^m  \left( k_0 a, \mathrm{cos} \vartheta'  \right)
	\nonumber\\[2mm]       &\times 
	R_{mn}^{(3)} \left( k_0 a, \mathrm{cosh} \xi^{\ge}  \right)
	R_{mn}^{(1)} \left( k_0 a, \mathrm{cosh} \xi^{\le}  \right)
	\mathrm{e}^{\mathrm{i} m \left( \varphi - \varphi' \right) } , 
\end{align}
and could also be expanded using spherical eigenfunctions into \cite{Tai_1971}
\begin{align}
	\label{eq:G0 appendix expansion 2}
	G_0 (\mathbf{r}, \mathbf{r}') &= \frac{\mathrm{i}k_0}{2\uppi}   \sum_{m,n}  (2-\delta_{m,0})    \overline{P}_n^m \left( \mathrm{cos}\theta \right)  \overline{P}_n^m \left( \mathrm{cos}\theta' \right)  
	\nonumber\\  &\times
	z_n^{(3)} \left(k_0r^{\ge}\right)  z_n^{(1)} \left( k_0r^{\le}\right)   
	\mathrm{e}^{\mathrm{i} m(\phi - \phi') }, 
\end{align}
where $\overline{\mathscr{P}}_n^m  \left( k_0 a, \mathrm{cos} \vartheta  \right) = N_{mn}^{-1} (k_0a) \mathscr{P} _n^m  \left( k_0 a, \mathrm{cos} \vartheta  \right)$, and $\overline{P}_l^m\left( \mathrm{cos} \theta \right) = N_{mn}^{-1} (0) P_l^m\left( \mathrm{cos} \theta \right)$. 
Moreover, $z_n^{(3)} (x) = h_n^{(1)} (x)$ is the spherical Bessel function of the third type, also called the spherical Hankel function, and $z_n^{(1)} (x) = j_n (x)$ is the spherical Bessel function of the first type.

By using the addition theorem of Legendre function \cite{WangZhuxi}
\begin{align}
	\label{eq:the addition theorem of Legendre functions 2nd version}
	P_n(\mathrm{cos} \Theta) = \frac{2}{2n+1} \sum_{m=-n}^{n}   
	\overline{P}_n^m \left( \mathrm{cos}\theta \right)  \overline{P}_n^m \left( \mathrm{cos}\theta' \right)
	\mathrm{e}^{\mathrm{i} m(\phi - \phi') } 
\end{align}
where
\begin{eqnarray}
	\label{eq:intersection angle}
	\mathrm{cos} \Theta  = \mathrm{cos} \theta  \mathrm{cos} \theta'  +  \mathrm{sin} \theta  \mathrm{sin} \theta' \mathrm{cos} \left( \phi - \phi'  \right) ,
\end{eqnarray}
and $\sqrt{2\uppi} \, Y_l^m (\theta, \phi) =  \overline{P}_l^m (\mathrm{cos} \theta) \mathrm{e}^{\mathrm{i} m \phi }$, we get
\begin{align}
	\label{eq:the addition theorem of Legendre functions 3rd version}
	2\uppi \sum_{k=-l}^{l}  Y_l^{k} \left( \theta , \phi  \right)  
	{Y_{l}^{k}}^* \left( \theta', \phi' \right)  
	= \frac{2n+1}{2}  P_n(\mathrm{cos} \Theta)   .
\end{align}

\end{appendices}

\bibliographystyle{unsrt}
\bibliography{sample}

\begin{thebibliography}{10}

\bibitem{Bohren2008}
Bohren~C F and Huffman~D R.
\newblock {\em Absorption and scattering of light by small particles}.
\newblock John Wiley \& Sons, 2008.

\bibitem{Wiscombe:80}
W.~J. Wiscombe.
\newblock Improved mie scattering algorithms.
\newblock {\em Appl. Opt.}, 19(9):1505--1509, May 1980.

\bibitem{Mishchenko2000}
M.~I. Mishchenko, J.~W. Hovenier, and L.~D. Travis.
\newblock {\em Light Scattering by Nonspherical Particles: Theory,
  Measurements, and Applications}.
\newblock IOP Publishing, 2000.

\bibitem{Mishchenko2002}
M.~I. Mishchenko, L.~D. Travis, and A.~A. Lacis.
\newblock {\em Scattering, absorption, and emission of light by small
  particles}.
\newblock Cambridge University Press, 2002.

\bibitem{Mishchenko2014}
M.~I. Mishchenko.
\newblock {\em Electromagnetic scattering by particles and particle groups: an
  introduction}.
\newblock Cambridge University Press, 2014.

\bibitem{Farafonov2013}
Victor Farafonov.
\newblock {\em Application of non-orthogonal bases in the theory of light
  scattering by spheroidal particles}, pages 189--266.
\newblock Springer Berlin Heidelberg, Berlin, Heidelberg, 2013.

\bibitem{Tai_1971}
Tai~C T.
\newblock {\em Dyadic Green's functions in electromagnetic theory}.
\newblock IEEE Press, 1971.

\bibitem{LiLewei2004}
Li~L W, Kang~X K, and Leong~M S.
\newblock {\em Spheroidal wave functions in electromagnetic theory}.
\newblock John Wiley \& Sons, 2004.

\bibitem{Akkermans2007}
E.~Akkermans and G.~Montambaux.
\newblock {\em Mesoscopic physics of electrons and photons}.
\newblock Cambridge University Press, 2007.

\bibitem{Carminati2021}
R.~Carminati and J.~C. Schotland.
\newblock {\em Principles of Scattering and Transport of Light}.
\newblock Cambridge University Press, 2021.

\bibitem{Mishchenko2006}
M.~I. Mishchenko, L.~D. Travis, and A.~A. Lacis.
\newblock {\em Multiple scattering of light by particles: radiative transfer
  and coherent backscattering}.
\newblock Cambridge University Press, 2006.

\bibitem{Piraud_2013}
M~Piraud, L~Pezzé, and L~Sanchez-Palencia.
\newblock Quantum transport of atomic matter waves in anisotropic
  two-dimensional and three-dimensional disorder.
\newblock {\em New Journal of Physics}, 15(7):075007, jul 2013.

\bibitem{Vollhardt1980}
D.~Vollhardt and P.~W\"olfle.
\newblock Diagrammatic, self-consistent treatment of the anderson localization
  problem in $d\ensuremath{\le}2$ dimensions.
\newblock {\em Phys. Rev. B}, 22:4666--4679, Nov 1980.

\bibitem{Tiggelen_2016}
Nicolas Cherroret, Dominique Delande, and Bart~A. van Tiggelen.
\newblock Induced dipole-dipole interactions in light diffusion from point
  dipoles.
\newblock {\em Phys. Rev. A}, 94:012702, Jul 2016.

\bibitem{Carminati2014}
Kevin Vynck, Romain Pierrat, and R\'emi Carminati.
\newblock Multiple scattering of polarized light in disordered media exhibiting
  short-range structural correlations.
\newblock {\em Phys. Rev. A}, 94:033851, Sep 2016.

\bibitem{Amic_1996}
E~Amic, J~M Luck, and Th~M Nieuwenhuizen.
\newblock Anisotropic multiple scattering in diffusive media.
\newblock {\em Journal of Physics A: Mathematical and General}, 29(16):4915,
  aug 1996.

\bibitem{Sheng2006}
P.~Sheng.
\newblock {\em Introduction to Wave Scattering, Localization and Mesoscopic
  Phenomena}.
\newblock Springer, 2006.

\bibitem{RevModPhys.71.313}
M.~C.~W. van Rossum and Th.~M. Nieuwenhuizen.
\newblock Multiple scattering of classical waves: microscopy, mesoscopy, and
  diffusion.
\newblock {\em Rev. Mod. Phys.}, 71:313--371, Jan 1999.

\bibitem{Tsang1980}
L.~Tsang and J.~A. Kong.
\newblock {Multiple scattering of electromagnetic waves by random distributions
  of discrete scatterers with coherent potential and quantum mechanical
  formalism}.
\newblock {\em Journal of Applied Physics}, 51(7):3465--3485, 07 1980.

\bibitem{Tsang1981}
L.~Tsang and J.~A. Kong.
\newblock {Multiple scattering of acoustic waves by random distributions of
  discrete scatterers with the use of quasicrystalline‐coherent potential
  approximation}.
\newblock {\em Journal of Applied Physics}, 52(9):5448--5458, 09 1981.

\bibitem{LiLewei2001}
Le-Wei Li, Mook-Seng Leong, Pang-Shyan Kooi, and Tat-Soon Yeo.
\newblock Spheroidal vector wave eigenfunction expansion of dyadic green's
  functions for a dielectric spheroid.
\newblock {\em IEEE Transactions on Antennas and Propagation}, 49(4):645--659,
  2001.

\bibitem{Farafonov_2016}
V.~G Farafonov, N.~V. Voshchinnikov, and E.~G. Semenova.
\newblock Some relations between the spheroidal and spherical wave functions.
\newblock {\em Journal of Mathematical SciencesJournal of Mathematical
  Sciences}, 214:382--391, 2016.

\bibitem{Varshalovich1988}
D~A Varshalovich, A~N Moskalev, and V~K Khersonskii.
\newblock {\em Quantum Theory of Angular Momentum}.
\newblock WORLD SCIENTIFIC, 1988.

\bibitem{WangZhuxi}
Z~X Wang and D~R Guo.
\newblock {\em Special Functions}.
\newblock WORLD SCIENTIFIC, 1989.

\bibitem{Rammer2004}
Jorgen Rammer.
\newblock {\em Quantum Transport Theory}.
\newblock CRC Press, 2004.

\bibitem{LAGENDIJK1996143}
Ad~Lagendijk and Bart~A. {van Tiggelen}.
\newblock Resonant multiple scattering of light.
\newblock {\em Physics Reports}, 270(3):143--215, 1996.

\bibitem{adelman2014softwarecomputingspheroidalwave}
Ross Adelman, Nail~A. Gumerov, and Ramani Duraiswami.
\newblock Software for computing the spheroidal wave functions using arbitrary
  precision arithmetic, 2014.

\bibitem{Stratton1941}
Adams~Julius Stratton.
\newblock {\em Electromagnetic Theory}.
\newblock Mcgraw Hill Book Company, 1941.

\bibitem{Table_of_integrals}
Gradshteyn~I S and Ryzhik~I M.
\newblock {\em Table of integrals, series, and products}.
\newblock Academic press, 2014.

\end{thebibliography}

\end{document}